\documentclass{sigchi}

\toappear{\scriptsize Permission to make digital or hard copies of all or part of this work for personal or classroom use is granted without fee provided that copies are not made or distributed for profit or commercial advantage and that copies bear this notice and the full citation on the first page. Copyrights for components of this work owned by others than ACM must be honored. Abstracting with credit is permitted. To copy otherwise, or republish, to post on servers or to redistribute to lists, requires prior specific permission and/or a fee. Request permissions from permissions@acm.org. \\
{\emph{CHI '20, April 25--30, 2020, Honolulu, HI, USA.} } \\
\copyright~2020 Association for Computing Machinery. \\
ACM ISBN 978-1-4503-6708-0/20/04\ ...\$15.00. \\
https://dx.doi.org/10.1145/3313831.3376553}

\usepackage{balance}       
\usepackage{graphics}      
\usepackage[T1]{fontenc}   
\usepackage{txfonts}
\usepackage{mathptmx}
\usepackage[pdflang={en-US},pdftex]{hyperref}
\usepackage{color}
\usepackage{booktabs}
\usepackage{textcomp}
\usepackage[utf8]{inputenc}

\usepackage{framed}
\usepackage{algorithm}
\usepackage[noend]{algorithmic}
\usepackage{graphicx}
\usepackage{multirow}
\usepackage{cite}

\usepackage{microtype}        
\usepackage{ccicons}          

\usepackage{xcolor}

\newcommand{\setofElements}{\ensuremath{ 		\mathbb{E} 	}~}
\newcommand{\someElement}{\ensuremath{		{e}			}}
\newcommand{\someOtherElement}{\ensuremath{	\bar{e}	}}

\newcommand{\WidthOfCanvas}{\ensuremath{ 			\mathbb{W} 		}~}
\newcommand{\HeightOfCanvas}{\ensuremath{ 			\mathbb{H} 		}~}

\newcommand{\LOfElement}[1]{\ensuremath{ 	{L}_{#1}}~}
\newcommand{\ROfElement}[1]{\ensuremath{ 	{R}_{#1}}~}
\newcommand{\TOfElement}[1]{\ensuremath{ 	{T}_{#1}}~}
\newcommand{\BOfElement}[1]{\ensuremath{ 	{B}_{#1}}~}
\newcommand{\WOfElement}[1]{\ensuremath{ 	{W}_{#1}}~}
\newcommand{\HOfElement}[1]{\ensuremath{ 	{H}_{#1}}~}

\newcommand{\ElementAboveElement}[2]{\ensuremath{\Gamma_{#2}^{#1}}}
\newcommand{\ElementBeforeElement}[2]{\ensuremath{\Pi_{#1 #2}}}

\newcommand{\toolname}{\textsc{Grids}}

\def\plaintitle{GRIDS: Interactive Layout Design with Integer Programming}
\def\plainauthor{Niraj Ramesh Dayama, Kashyap Todi, Taru Saarelainen, Antti Oulasvirta}

\def\plainkeywords{Grid Layouts; Creativity Support; Computational Design; Mixed-Initiative; Optimisation; Design Tools}

\makeatletter
\def\url@leostyle{%
  \@ifundefined{selectfont}{
    \def\UrlFont{\sf}
  }{
    \def\UrlFont{\small\bf\ttfamily}
  }}
\makeatother
\def\pprw{8.5in}
\def\pprh{11in}

\newcommand{\compresslist}
{
    \setlength{\itemsep}{1pt}
    \setlength{\parskip}{0pt}
    \setlength{\parsep}{0pt}
}

\definecolor{linkColor}{RGB}{6,125,233}
\hypersetup{%
  pdftitle={\plaintitle},
  pdfauthor={\plainauthor},
  pdfkeywords={\plainkeywords},
  pdfdisplaydoctitle=true, 
  bookmarksnumbered,
  pdfstartview={FitH},
  colorlinks,
  citecolor=black,
  filecolor=black,
  linkcolor=black,
  urlcolor=linkColor,
  breaklinks=true,
  hypertexnames=false
}

\usepackage{cuted}
\usepackage{capt-of}

\begin{document}

\title{\plaintitle}

\author{
\alignauthor{Niraj Ramesh Dayama\thanks{Authors contributed equally}*\textsuperscript{1}\hspace{1cm} Kashyap Todi*\textsuperscript{1,2}\hspace{1cm} Taru Saarelainen\textsuperscript{1} \hspace{1cm} Antti Oulasvirta\textsuperscript{1,2}\\
\affaddr{\textsuperscript{1}Department of Communication and Networking,}
\affaddr{Aalto University, Finland}\\
\affaddr{\textsuperscript{2}Finnish Center for Artificial Intelligence FCAI, Finland}\\
\email{niraj.dayama@aalto.fi},
\email{kashyap.todi@gmail.com},
\email{saarelainen.taru@gmail.com},
\email{antti.oulasvirta@aalto.fi}
}
}



\teaser{
\centering
\includegraphics[width=0.98\textwidth]{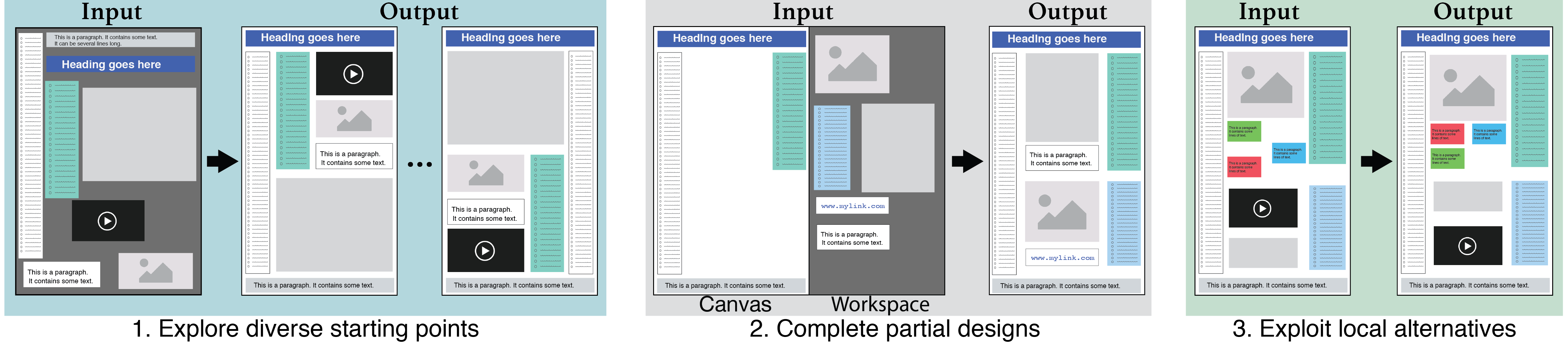}
\caption{This paper presents an integer programming approach for interactively generating grid layouts, enabling designers to (1) systematically explore a diverse set of starting solutions, (2) find solutions to partially-completed designs, and (3) search for local alternatives and within sub-spaces.  }
\label{fig:teaser}
}

\maketitle



\begin{abstract}

Grid layouts are used by designers to spatially organise user interfaces when sketching and wireframing. 
However, their design is largely time consuming manual work.
This is challenging due to combinatorial explosion and complex objectives, 
such as alignment, balance, and expectations regarding positions.
This paper proposes a novel optimisation approach for the generation of diverse grid-based layouts. 
Our mixed integer linear programming (MILP) model offers a rigorous yet efficient method for grid generation that ensures packing, alignment, grouping, and preferential positioning of elements.
Further, we present techniques for interactive diversification, enhancement, and completion of grid layouts (\autoref{fig:teaser}).
These capabilities are demonstrated using \toolname\footnote{\textbf{\emph{GRIDS}}: \textbf{\underline{G}}enerating \textbf{\underline{R}}eal-time \textbf{\underline{I}}nterface \textbf{\underline{D}}esign \textbf{\underline{S}}uggestions}, a wireframing tool that provides designers with real-time layout suggestions.
We report findings from a ratings study ($N=13$) and a design study ($N=16$), lending evidence for the benefit of computational grid generation during early stages of design.

\end{abstract}



\section{Introduction}
\label{Introduction}
\begin{figure}
\centering
\includegraphics[width=0.95\columnwidth]{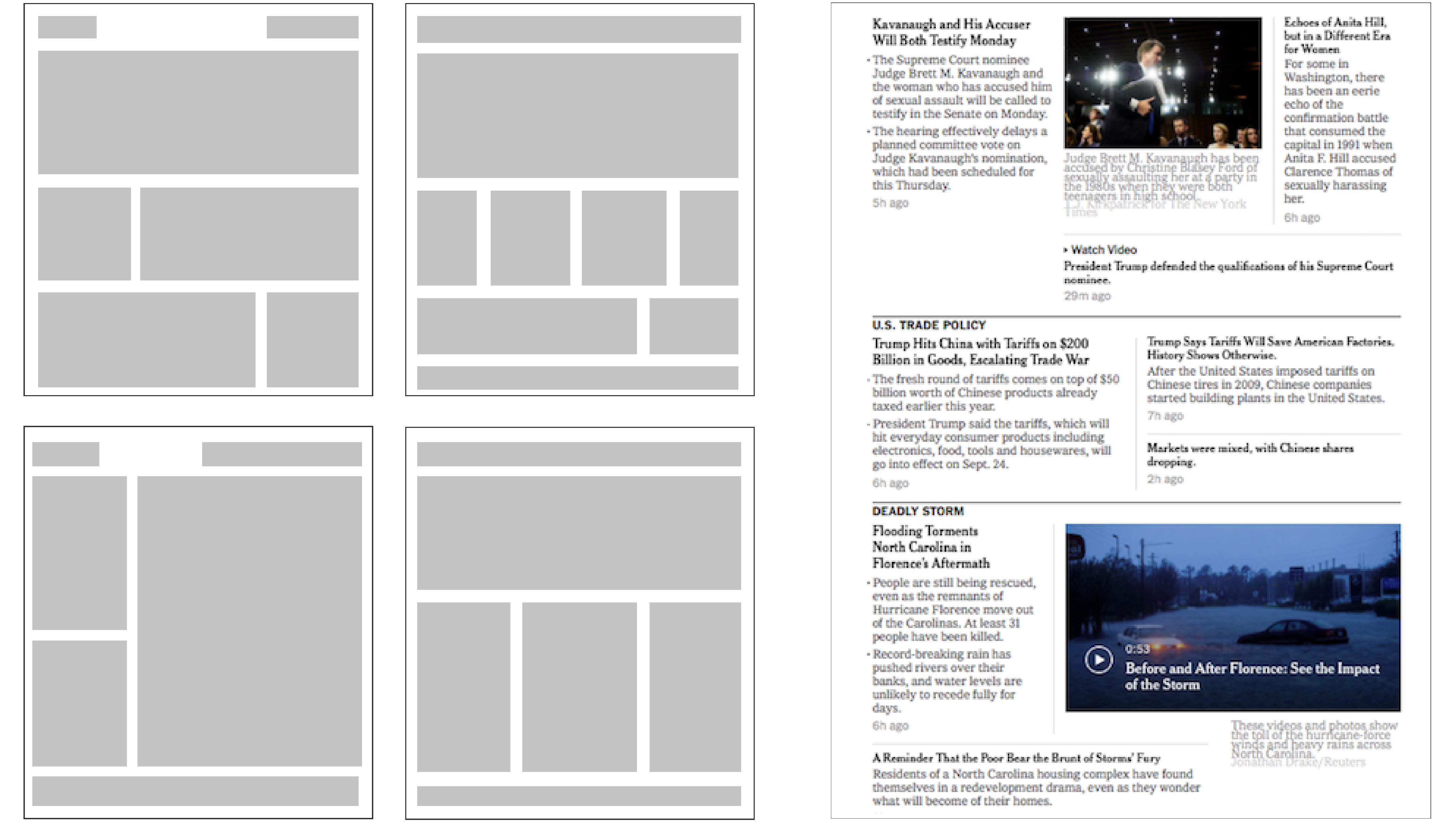}
\caption{Grid layouts are used to organise visual elements in different stages of design. Left: Examples of grid layout templates; Right: A Web interface following a CSS-based grid template (\url{www.nytimes.com}).}
\label{fig:gridlayouts}
\end{figure}

The design of graphical user interfaces (GUIs) commonly employs \emph{grid layouts} \cite{lupton2014thinking} -- spatial structures defined by grid lines.
Grid lines simplify layout design by guiding the sizes and positions of GUI elements.
Grids serve as a mechanism for evaluating and deciding their spatial organisation,
including considerations of visual importance, grouping, relative positioning, and flow.
Grid layouts are used across various stages of design, from sketching and wireframing to prototyping and deployment (\autoref{fig:gridlayouts}).

The design of grid-based layouts is mostly manual and left to the designer.
Although present-day tools provide functionalities such as grid templates, grid-snapping, and usability hints \cite{lynch2008web},
they mostly do not offer computational layout generation.
Editing involves repetitive manual work including resizing and reorganising.
Design is also combinatorially challenging.
There exists a large number of possibilities for organising a given set of elements into a grid,
and a good grid layout must meet several objectives \cite{koffka_principles_2013,lok_evaluation_2004}.
Changing one element may necessitate reorganisation of many others.

Computational generation of grid layouts has the potential to support designers in the creative process of layout design,
helping them synthesise, envision, and evaluate \cite{philippaontology}.
Access to diverse but relevant computationally generated suggestions could help them in exploring and enhancing designs \cite{todi_sketchplore:_2016}, thus avoiding too early fixation \cite{chan2017semantically}.
Moreover, need for manual editing could be reduced if computer-generated designs were able to suggest completions to partially finalised designs.
However, the algorithmic problems involved are non-trivial.
Any such method must efficiently search within a very vast design space  while reacting to real-time changes made by the designer.
Prior to this work, it has not been known how to best define core features of a \emph{good} grid layout.
The more design objectives and concerns one can incorporate into the mathematical system, the less final editing is left to the designer.
However, it is not known how to combine objectives that together ensure performant and aesthetically pleasing layouts.
Elements must be well-aligned, key elements easily reachable, related elements logically grouped together, and so on.

We propose a mathematical formulation of the grid layout problem that enables real-time generation and exploration of layout designs with a designer in the loop.
Our approach is based on \emph{integer programming} (IP) \cite{wolsey1998integer}.
One benefit of IP---a so-called \emph{exact} method---over random-search (black-box) based approaches is that the key features of the expected outputs can be \emph{guaranteed}.
If a solution exists, it will be found.
Further, we use a combination of continuous and discrete decision variables, thus resulting in a \emph{mixed integer linear programming} (MILP) formulation.
In our model, \emph{continuous} decision variables are used for element placement and sizing, and \emph{discrete} decision variables are used to define the relative positions of elements with reference to each other.
Doing so makes the model independent of canvas-size and element-size, drastically reducing the number of discrete variables required.
Our formulation strictly uses \emph{linear} expressions (all constraints and in objectives), making the model computationally fast and enabling interactive use even for larger-sized problems.

Our MILP model guarantees \emph{proper packing}:
all elements fit on the canvas without overflowing or overlapping.
In a single model, it further addresses other design goals such as:
\textbf{(1)} the outer hull is rectangular,
\textbf{(2)} there are no holes,
\textbf{(3)} elements are well-aligned \cite{miniukovich2015computation},
\textbf{(4)} related elements are grouped together, and
\textbf{(5)} preferred positions are obeyed to provide visual connectivity \cite{balinsky_aesthetic_2009}.
Prior works on computational grid generation \cite{sears_aide:_1995, balinsky_aesthetically-driven_2009} have mostly addressed one or few of these key properties at a time.
As IP provides \emph{bounds} for its solutions, designs can be guaranteed to be within specified range (say within 5\%) from the best achievable design.
We exploit this for generation of \emph{controllably diverse designs}.

Our approach offers several benefits as a computational method.
First, many necessary layout properties are included in the formulation,
reducing effort for the designer.
Second, making the model canvas-size independent results in a direct advantage.
Previously-employed approaches (e.g. \cite{todi_sketchplore:_2016}) mostly relied on discretisation of the canvas.
For example, a problem of 5 elements placed on an 800$\times$600 pixel canvas would require around 2.4 million discrete decision variables and constraints to ensure a non-overlapping and well-aligned layout.
For the same 5-element problem, our MILP model requires only 110 discrete and 20 continuous decision variables.
Such problems can now be solved to generate high quality grid layout solutions within few seconds using commodity hardware.
Finally, black box methods for multi-objective optimisation often yield an unbalanced sample from the so-called Pareto front.
In other words, many solutions may be too similar with each other to be informative.
In contrast, we build on the exact MILP formulation to guarantee \emph{diverse solutions} that are individually well-performing -- that is, close to optimal for all objectives -- and space-spanning, i.e. diverse in a systematic way.
This enables designers to explore good alternatives more efficiently.

These properties support integration within interactive design tools.
To demonstrate this, we present \toolname, an interactive wireframing tool that encapsulates our MILP approach to present real-time design suggestions.
Designers can use \toolname\ for exploring and improving grid layouts with little effort.
As illustrated in \autoref{fig:teaser}, auto-generated suggestions help designers to interactively (1) explore a diverse set of starting points towards solving a layout problem, (2) find completed solutions for partially-designed layouts, (3) exploit nearby alternatives and solutions within a specified subspace of the design space.
All suggestions are presented in an interactive example gallery \cite{Lee:2010:DIE:1753326.1753667},
with the idea that designers quickly recognise good designs (see \cite{Tanner2019}).
We assess these capabilities empirically with professional designers,
reporting quantitative and qualitative feedback.
The results indicate that this approach can assist designers in the exploration of GUI layouts in realistic design tasks.

To sum up, the key contributions of this paper are:
\begin{enumerate}\compresslist
    \item A fast and comprehensive mixed-integer linear programming (MILP) model for grid layout generation.
    \item Further computations for diversification, intensification, and completion of partial solutions.
    \item Demonstration of interactive grid generation in the \toolname\ tool, and evaluation with 16 professional designers.
\end{enumerate}

\section{Related Work}\label{relatedwork}

The grid layout is a spatial organisational principle \cite{hurlburt_grid:_1982}
developed post-WWII \cite{muller-brockmann_grid_1996}. 
We here focus on rectangular grids typical in the design of graphical user interfaces (GUIs).

\subsection{Facilitating the creation of grid layouts}

Grid layouts are utilised as a design principle in many toolkits and layout managers \cite{lok_survey_2001}, which  offer interactive aids like grid-snapping and auto-alignment.
A ruler-and-compass metaphor has been used \cite{bier_snap-dragging_1986} to aid in precise placement.
Here, heuristics to automatically place guiding lines and circles were used, helping users construct grid-adhering shapes.
Interactive grids and multi-touch alignment guides have also been presented to guide grid positioning \cite{frisch_grids_2011}.
A method for global beautification has been presented that infers relationships between layout elements, and fixing issues such as misalignment \cite{xu_global_2014}.
Users can interactively refine the layout by resolving ambiguity and adding constraints.
While these techniques assist with grid design and alignment,
layout construction is left to the designer.
Design tools also often offer grid templates: predefined layouts where content can be added \cite{drenttel_method_1999}.
However, templates insist that contents are made to fit them,
and they are limited to a small set of pre-defined layouts.

\subsection{Grid generation by constraint solving}

Since its first introduction in Sketchpad \cite{sutherland_sketchpad:_1963}, constraints that can aid in the design of layouts has been a topic of research \cite{borning_constraints_1997,oney_constraintjs:_2012,gleicher_briar:_1992, 10.1145/882262.882353}.
Layout constraints can define bounds on elements, or relationship between elements.
A constraint solver manipulates element properties to best satisfy the specified constraints.
One method is to solve layout constraints incrementally \cite{freeman-benson_incremental_1990}.
It can satisfy changing constraint hierarchies consisting of hard and soft constraints in an efficient manner.
Relational grammars can be further added to encode design knowledge \cite{weitzman_automatic_1994}.
This can address the logical composition of elements during layout generation.
Chorus \cite{hosobe_modular_2001} addressed non-linear geometric constraints such as Euclidean geometric, non-overlapping, and graph layout constraints.
It also discussed soft constraints with hierarchical strengths or preferences.
Cassowary \cite{badros_cassowary_2001} implemented a linear arithmetic constraint solver to adapt layout to changing sizes.
This has also been implemented in some commercial systems such as Apple's Auto Layout\footnote{https://overconstrained.io}, enabling GUIs to dynamically adjust their layouts when window or screen dimensions changed.
A benefit of constraint solving is that a layout can be defined by means of simple constraints.
On the other hand, specifying a layout fully using just constraints quickly becomes complex \cite{zeidler_comparing_2012}.
To our knowledge, no constraints-based method has been proposed that ensures proper packing of elements and takes care of objectives like alignment, rectangularity, and grouping.
Lacking this, designers may need to modify layouts manually or fix remaining aspects programmatically.

\subsection{Grid generation by combinatorial optimisation}

Combinatorial optimisation methods can find solutions, from a large design space, that satisfy a stated design objectives.
Within the area of combinatorial geometry,
grid layout has been studied in the context of 2D bin packing \cite{LODI2002241,doi:10.1287/mnsc.44.3.388, PISINGER2005154}, rectangular packing \cite{HART1995244}, and the guillotine cuts problem \cite{CHRISTOFIDES199521}. 
An elementary version of the grid design problem has been previously proposed \cite{HART1995244}, but it merely attempts to find the most densely packed solution whereby the elements are squeezed together as closely as possible; there is no attempt to address aesthetics (like alignment) and no intent to explore multiple layouts.   
Optimisation has been used for resizing GUI elements \cite{zeidler_constraint_2012}, with the focus being on improving aesthetics of a layout by making subtle changes when the available space is varied.
Previous research has also looked into generating optimised grid-based layouts.
AIDE is a metrics-based tool to aid design and evaluation of UIs \cite{sears_aide:_1995}.
It included objectives such as balance, efficiency, and constraints.
The optimiser computed a layout grid and used this to organise widgets.
ADL is a layout engine that dynamically constructed grid layouts \cite{balinsky_aesthetically-driven_2009}.
Given static elements, it filled up the remaining area with dynamic content
.
Typically, these approaches compute a single point-optimal solution.
Following earlier findings, we believe it is necessary to present designers with a diverse range of solutions \cite{bailly_menuoptimizer:_2013,todi_sketchplore:_2016}.
However, random-search based optimisation methods do not support generation of controllably diverse results.
IP-based approaches have been previously used to generate, for example, keyboard layouts
\cite{karrenbauer_improvements_2014}.
This allows for a structured search process and guaranteed bounds, thus improving outcomes and increasing designer confidence. 
More recently, AdaM employed IP to distribute UIs in multi-user environments \cite{Adam}. 
While the formulation does address limited display size, the objectives, constraints, and applications differ substantially from ours.   

SUPPLE and ARNAULD constitute the seminal literature concerning personalised UI generation \cite{ArnaultSource,gajos_supple:_2004,SUPPLE2,SUPPLE3,SUPPLEPLUSPLUS,CompareSupple}. 
The systems automatically generate interfaces through a branch-and-bound optimisation technique that is flexible enough to adapt to various objectives by modifying the underlying cost function. 
Our work differs from this line of work in three key aspects:
(1) \emph{Input}: SUPPLE and its descendants require functional specifications as input.
This is beneficial for application developers as they only need to specify what functions are exposed to users.
On the contrary, our work targets UI designers who already have some ideas about desired layouts.
It enables designers to define the design task by demonstration, including element types, and preferences regarding size and placement. 
(2) \emph{Layouting}: When rendering an interface layout, SUPPLE used three heuristics to sequentially decide the layout -- bottom-up, top-down, and minimum remaining values \cite{gajos2008automatically}. 
The key objective was to find a layout that satisfies user preferences, abilities, and device constraints. 
In contrast, our MILP formulation aims at constructing well-formed grid layouts, with desirable aesthetics, by considering the entire composition of the canvas. 
(3) \emph{Output}: SUPPLE and others aim at producing a single point-optimal design for end-users.
In contrast, our MILP approach generates multiple diverse solutions for mixed-initiative design tools. 


\subsection{Interactive layout generation}

Computational techniques have been proposed for designing layouts with designer in the loop.
DesignScape \cite{odonovan_designscape:_2015} supports enhancement and exploration of single-page layouts using energy-based optimisation.
The design principles here are limited to those learnt from a small set of examples, and optimisation is computationally-expensive.
Sketchplorer \cite{todi_sketchplore:_2016} supports exploring, enhancing, and recolouring layouts. 
However, it employs black-box optimisation on a discrete canvas, which does not scale, resulting in poor performance for complex design tasks. 
Moreover, it does not address objectives of good grid layouts, such as rectangularity, alignment, or preferential placement. 
Genetic algorithms have been used to iteratively select promising interfaces from a collection of candidates  \cite{du_plessis_incorporating_2008}.
Various constraints had to be placed to prevent inappropriate groupings in the layout, and to reduce the number of selections.
Data-driven approaches enable layout generation without requiring problem specification \cite{zheng2019content, DBLP:conf/iclr/LiYHZX19}.
However, results are limited to the domain of the training dataset, and offer no quality guarantees.
In what comes close to our aims, a semi-automatic grid-based technique has been previously proposed \cite{hudson_synergistic_1993}.
Here, users constructed an example layout, which, as in this paper, was represented as an algorithmic task to generate layouts of any size.
New examples were presented to the user, enabling them to steer the generation.
Interactive genetic algorithms have also been used \cite{Quiroz:2007:HGE:1240866.1241052} to generate grid layouts that address contrast between elements, and some alignment issues.
However, aspects such as rectangularity, and preferential placement were not addressed, thus leaving these aspects to the designer. 
Our work proposes a more complete formulation that takes into account these key aspects of a well-formed grid layout.




\subsection{Summary}

Previous works have studied aspects of grid-based layouts individually, and provided point solutions to some features (packing, alignment, etc).
This paper is the first attempt to combine key objectives within one compact formulation while guaranteeing the quality of layouts, providing diverse space-spanning solutions, and real-time performance.


\section{The Grid Layout Problem}
\label{sec:gridproblem}
\begin{figure}[t!]
\centering
\includegraphics[width=0.99\columnwidth]{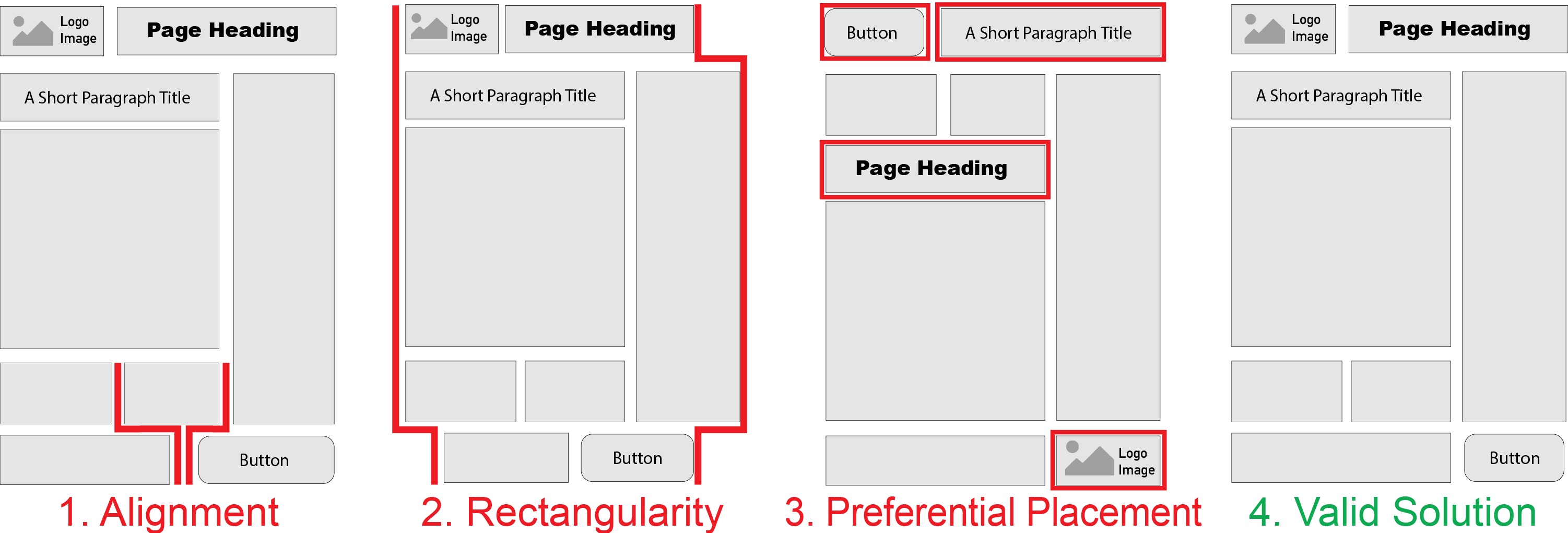}
\caption{Key objectives in the grid layout formulation. 
(1) \emph{Alignment}: red lines indicate misalignment issues;
(2) \emph{Rectangularity}: red lines indicate a non-rectangular outline of the grid; 
(3) \emph{Preferential placement}: red borders highlight elements not placed at preferred positions. 
(4) A valid solution that conforms to all objectives.}
\label{fig:badgrids}
\end{figure}

The computational problem in generating grid layouts can be defined as follows: 
\begin{enumerate}
    \item [] \textit{Given a set of rectangular elements that must be placed onto a fixed size canvas with specified widths and heights, find feasible solutions where
all elements are properly packed in a non-overlapping and non-overflowing layout.}
\end{enumerate}
In our work, we formulate the following objectives (\autoref{fig:badgrids}):
\begin{enumerate}\compresslist
\item \textbf{Overall alignment}: A well-formatted layout places as many/most of its elements aligned to each other edge-wise.
\item \textbf{Rectangular outline}: An outline with jagged edges, a lop-sided hull, or any non-convex arrangement, is aesthetically undesirable. 
The overall outline induced by the layout of elements must approach a rectangular external hull.
\item \textbf{Placement}: We prefer to place interrelated elements in close proximity to each other. This transition objective manifests itself in three ways:
\begin{enumerate}
\item \textit{Traversal Distance}: If users often need to navigate between a pair of UI elements (for example, a text box and an associated button), it is preferable that the distance between these elements be minimised.
\item \textit{Grouping}: 
A contiguous placement should be ensured for semantically or otherwise related items.
\item \textit{Preferential Placement}: A designer may want a specific element be placed definitely on a specific \emph{side} of some other element, or at a specific point (\emph{locked}) on the canvas. 
For example, the `Search' button may be expected to be on the right side of a text entry field.
\end{enumerate}
\end{enumerate}
For \emph{interactive} design, in addition to the above objectives, we cover techniques to:
(1) systematically diversify designs, 
(2) respect pre-defined canvas elements when completing partial designs, and
(3) search locally for enhancements.

The following section provides the core MILP formulation, along with specific formulations to address each of the objectives.

\section{Mixed Integer Linear Programming (MILP) Model} \label{MIP Formulation} 

In this section, we highlight key mathematical concepts that enable us to generate solutions for the defined grid layout problem.
The complete mathematical formulation is included in the Supplementary Material.

\textbf{Overview:} We propose an optimisation model whose size depends on the number of elements alone; the number of decision variables and constraints in our model is independent of the canvas and other factors. 
Further, 
we make an explicit distinction between the (1) the \emph{core MILP formulation}, which only generates properly packed grid skeletons, and (2) the \emph{other objectives}. The core MILP assures a non-overlapping non-overflowing grid with elements placed within permissible size limits in permissible locations.  These candidate solutions from the core MILP may not be good, well-aligned or aesthetically acceptable. Rather, the other requirements of alignment, external outline, etc. are enforced over the core MILP skeleton by plugging in as objective functions or as additional constraints.

\subsection{Core MILP formulation for grid skeletons}
We define our coordinate system such that the origin/reference point is the top left corner of the canvas. The X axis is positive rightwards and the Y axis is positive downwards. Continuous decision variables $\LOfElement{\someElement}, \ROfElement{\someElement}, \TOfElement{\someElement}, \BOfElement{\someElement}$ represent the location of the left, right, top and bottom edges of individual element \someElement. 
Decision variables $\WOfElement{\someElement},\HOfElement{\someElement}$ represent the actual width and height. These variables ensure proper sizing and prevent overflow. 
We note that the variables \LOfElement{}, \ROfElement{}, \TOfElement{},
\BOfElement{} are not binary and do not involve discretisation of pixels or locations. This factor is critical for the performance of the solver.

Next, we prevent overlap using the following two \emph{binary} decision variables adapted from an MILP approach \cite{HART1995244} that was originally introduced for the rectangular packing problem:\\
$
\ElementAboveElement{\someElement}{\someOtherElement} \ \  \longrightarrow \ \  \mbox{Indicates \someElement~ is placed anywhere above \someOtherElement}\\
\ElementBeforeElement{\someElement}{\someOtherElement} \ \ \longrightarrow \ \  \mbox{Indicates \someElement~ is placed to the left side of \someOtherElement}
$

The interpretation of \ElementAboveElement{\someElement}{\someOtherElement} is
that the bottom edge \BOfElement{\someElement} of element \someElement~ is
less than or equal to the top edge \TOfElement{\someOtherElement} of element
\someOtherElement.
Similarly, the interpretation of \ElementBeforeElement{\someElement}{\someOtherElement} is
that the right edge \ROfElement{\someElement} of element \someElement~ is less
than or equal to the left edge \LOfElement{\someOtherElement} of element
\someOtherElement. Either of these decision variables
\ElementAboveElement{}{} and \ElementBeforeElement{}{} effectively slice the space
around any element into two half-spaces. 
Overlap avoidance practically means
that any other element must completely lie within one of those half-spaces.
Then the variables \ElementAboveElement{}{} and \ElementBeforeElement{}{} are sufficient
to prevent overlap of any pair of elements via the following constraints:
\begin{align}
&1 \le \ElementAboveElement{\someElement}{\someOtherElement} +
\ElementAboveElement{\someOtherElement}{\someElement}+\ElementBeforeElement{\someElement}{\someOtherElement}
+ \ElementBeforeElement{\someOtherElement}{\someElement} \le 2\ \dots\forall
\someElement, \someOtherElement \in \setofElements \label{MainNonOverlapEquation}\\
&\TOfElement{\someOtherElement} \ge \BOfElement{\someElement} +
\HeightOfCanvas(\ElementAboveElement{\someElement}{\someOtherElement} -1) \
\dots\forall \someElement, \someOtherElement \in \setofElements \label{FirstPlumbingEquation} \\
&\LOfElement{\someOtherElement} \ge \ROfElement{\someElement} +
\WidthOfCanvas(\ElementBeforeElement{\someElement}{\someOtherElement} -1) \
\dots\forall \someElement, \someOtherElement \in \setofElements\\
&\WidthOfCanvas \ElementBeforeElement{\someElement}{\someOtherElement} \ge
 \LOfElement{\someOtherElement} - \ROfElement{\someElement} \
\dots\forall \someElement, \someOtherElement \in \setofElements\\
&\HeightOfCanvas \ElementAboveElement{\someElement}{\someOtherElement} \ge
 \TOfElement{\someOtherElement} - \BOfElement{\someElement}  \
\dots\forall \someElement, \someOtherElement \in \setofElements
\label{LastPlumbingEquation}
\end{align}
Equations \eqref{FirstPlumbingEquation}--\eqref{LastPlumbingEquation} are `plumbing' constraints that logically connect the element edge locations to the \ElementBeforeElement{}{} and \ElementAboveElement{}{} variables.
There are four possible cases for two non-overlapping elements 
\someElement\ and \someOtherElement: (i) \someElement\ above \someOtherElement, (ii) \someElement\ below \someOtherElement, (iii) \someElement\ left of \someOtherElement, or (iv) \someElement\ right of \someOtherElement.
Equation \eqref{MainNonOverlapEquation} is the crux of the overlap prevention, which ensures that for any pair of elements, at least one of these cases is satisfied. 
This core formulation yields non-overflowing non-overlapping layouts with element sizes within their prescribed limits. 

\subsection{Ensuring overall alignment}
To model overall alignment, we define notional Cartesian grid-lines over all pixels of the canvas.
For any feasible solution, only those grid-lines are relevant where at least one element edge occurs. If two or more elements have any of their edges aligned with each other, those elements share a single grid-line for those edges.
So, the total number of grid-lines \textit{actually utilised} in any feasible solution quantifies the overall alignment within that solution. Poorly aligned feasible solutions require a larger number of grid-lines. 
Designers prefer well-aligned solutions and this objective translates to minimisation of the total number of grid-lines required.

Consider a candidate solution where one edge of element \someElement~ and
the corresponding edge of element \someOtherElement~ are aligned. Then we designate set \{\someElement,\someOtherElement\} as an
\textit{Aligned Group}. Formally, we define an \textit{Aligned Group} as a set
of elements whose one corresponding edge is aligned for the given candidate
solution. An aligned group may be a singleton (containing only one element) or may contain
all elements from \setofElements. The four distinct types of
aligned groups are:
\begin{enumerate}
    \item []  If the left edge of elements \someElement,
\someOtherElement, \dots are aligned, then we designate set
\{\someElement,\someOtherElement,\dots\} as a \textit{Left-Group}
abbreviated as \textit{LG}.
    \item [] If the right edge of elements \someElement,
\someOtherElement, \dots are aligned, then we designate set
\{\someElement,\someOtherElement,\dots\} as a \textit{Right-Group}
abbreviated as \textit{RG}.
    \item [] If the top edge of elements \someElement,
\someOtherElement, \dots are aligned, then we designate set
\{\someElement,\someOtherElement,\dots\} as a \textit{Top-Group}
abbreviated as \textit{TG}.
    \item [] If the bottom edge of elements \someElement,
\someOtherElement, \dots are aligned, then set
\{\someElement,\someOtherElement,\dots\} is designated as \textit{Bottom-Group}
abbreviated as \textit{BG}.
\end{enumerate}

\begin{figure}[t!]
  \includegraphics[width=0.95\columnwidth]{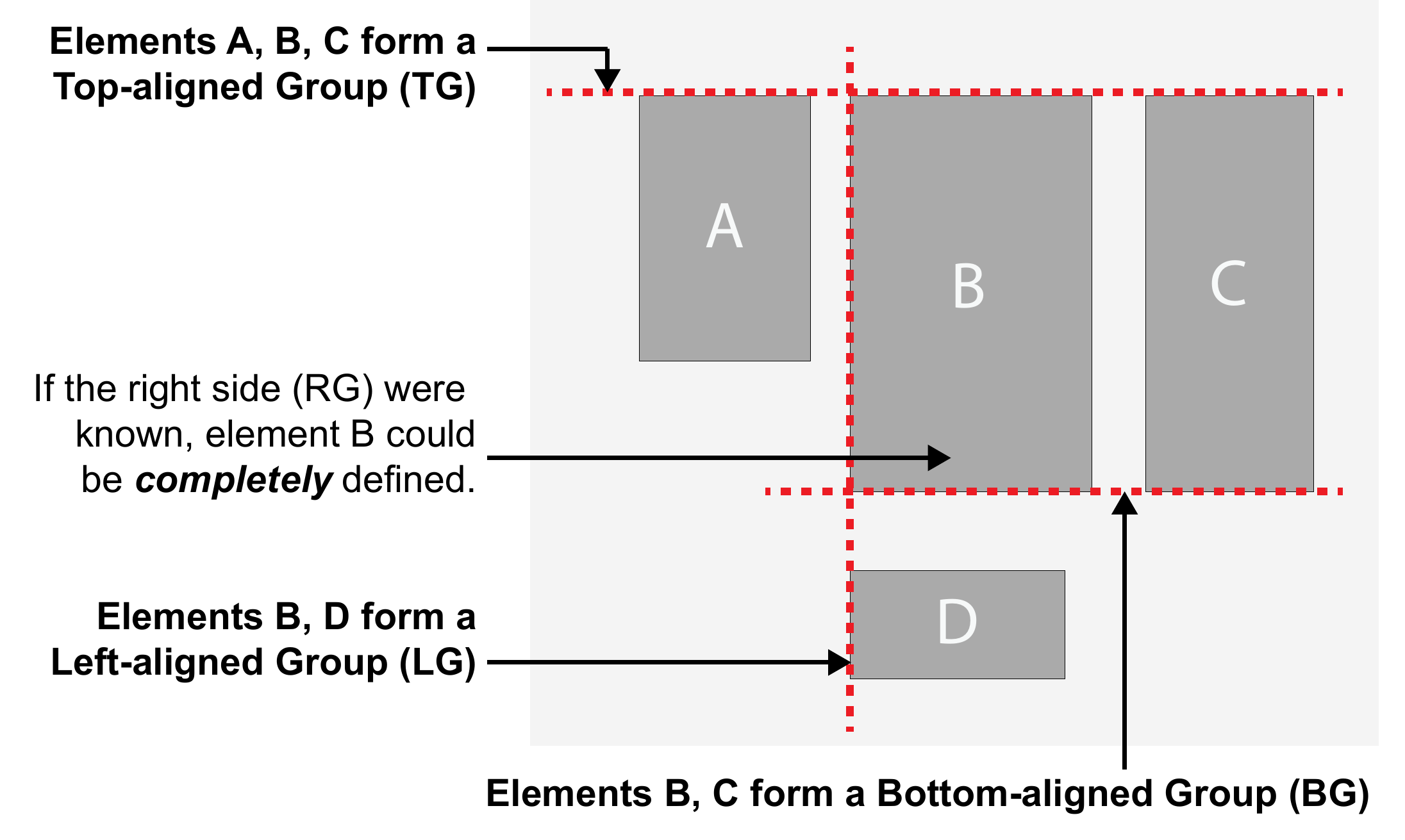}
  \caption{Alignment groups specify relative placement of groups of elements. An element is completely defined if it belongs to four such groups.}
  \label{AlignmentGroups}
\end{figure}
For every alignment-group, we define the \textit{value} as 
  the distance of the relevant edge of all its constituent elements from the
  corresponding axis. Decision variable $V^i$ indicates value of the $i^{th}$ alignment group. 
  Every element \someElement~ belongs to exactly one LG, one RG, one TG and
one BG. The four \textit{values} of these four alignment-groups match the edge
locations \LOfElement{\someElement}, \ROfElement{\someElement},
\TOfElement{\someElement} and \BOfElement{\someElement} of element
\someElement. So, the four alignment-groups of any element are
sufficient to define the location and size of that element. 
The overall layout is completely defined if the alignment-groups information is known for all elements.
The solver decides which alignment-groups are required in the solutions, 
the value of each group, and membership information (elements belonging to that group). 
The concept of alignment-groups is illustrated in Figure \ref{AlignmentGroups}.

\subsection{Ensuring rectangular outline}
Consider a layout of $n$ elements involving $m$ number of LGs.
No two LGs can share the same left-side value (otherwise they will become one single LG).
So, we can identify the specific LG that has the minimum left-most value.
This LG identifies the extreme-left side of the outline hull of the overall layout.
Similarly, we identify the right-most, top-most and bottom-most alignment groups.
Together, these extremities define the smallest rectangular outline (SRO) that would have covered the layout. In general, this SRO is not the immediate boundary of the layout. The actual boundary of the layout is a slightly smaller (potentially concave) shape within this SRO.
The difference between the true outline and the SRO is what makes the outline non-rectangular.
We explore several options to minimise and penalise this undesirable difference between the true outline and the SRO:
\begin{enumerate}\compresslist
\item \textbf{Adherence to extremity}: We identify the alignment-groups that define the SRO. The elements aligned to the SRO do not disturb the rectangularity. So, our objective function rewards every \textit{case} where an edge fits the SRO.
\item \textbf{Penalise non-extremity}: Consider an element $x$ whose right-most edge is at distance $D$ from the right-side of SRO.
Further, there is no element $y$ to the right-side of element $x$.
We ascribe the gap $D$ to element $x$ alone and penalise element $x$ proportional to gap $D$.
\item \textbf{Virtual elements}: We define virtual elements to fill any non-rectangular area (to be penalised). We penalise the existence of such virtual elements. This approach involves non-linearity and carries higher computational cost.
\end{enumerate}

The first approach is sufficient for the grid layout problem. We compute the maximum possible number of \textit{cases} where an edge of an element aligns with the extreme edges.
Thereafter, we ensure that the actual layouts generated must continue with (a slightly loosened approximation of) this maximum number.

\subsection{Ensuring preferential placement of elements}
A pair of elements \someElement~ and \someOtherElement~ can be placed \emph{relative to each other} using \ElementBeforeElement{}{} and \ElementAboveElement{}{} decision variables.
Elements can also be placed \emph{relative to the canvas}.
For example, consider that element \someElement~ is marked to be the header of the canvas.
This means that \someElement~ must necessarily lie \emph{above} all other elements.
In some cases, it is permissible that \someOtherElement ~ may be placed parallel to (but not above) the header.
All such conditions can be enabled by enforcing that no element \someOtherElement~ may ever be placed above the header \someElement.
Finally, elements can also have \emph{fixed positions} on the canvas using similar mechanisms.

\subsection{Generating controllably diverse solutions}
\begin{figure}
\centering
\includegraphics[width=0.9\columnwidth]{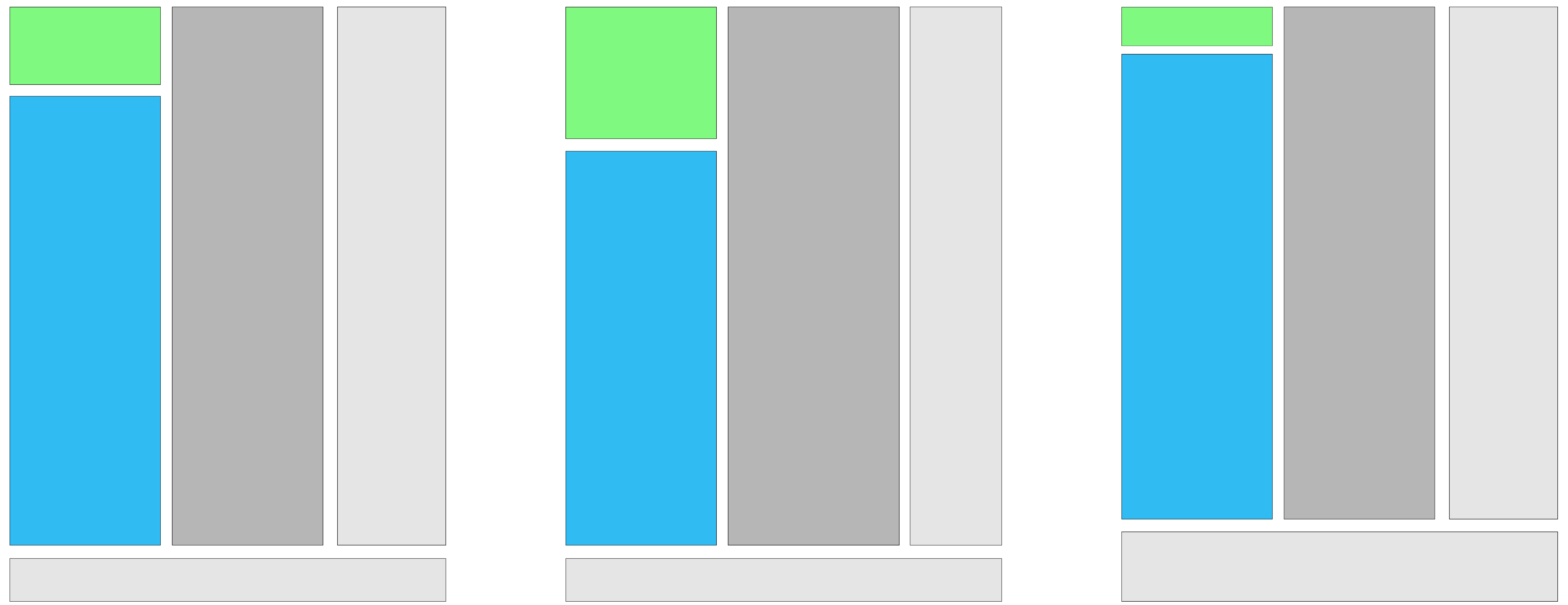}
\caption{When diversity is \emph{not} controlled, similar designs are generated.}
\label{fig:similargrids}
\end{figure}

While there has been some work in generating diverse layouts \cite{odonovan_designscape:_2015,todi_sketchplore:_2016}, it has been hard to accurately estimate how different designs are, with respect to each other.
\autoref{fig:similargrids} illustrates this problem. It shows some very similar solutions produced because the design space is not systematically spanned. 

\begin{figure}
  \includegraphics[width=0.9\columnwidth]{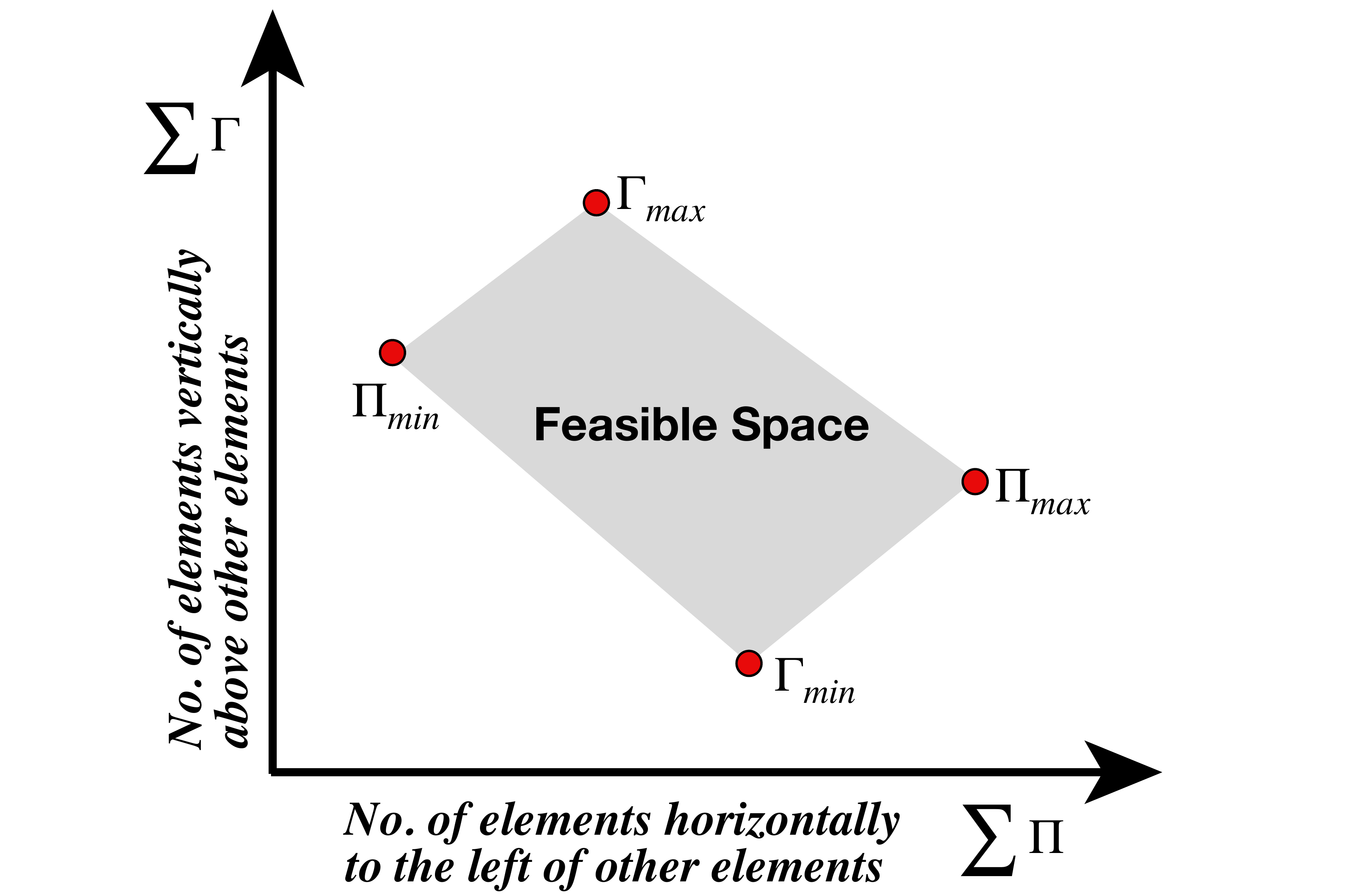}
  \caption{The feasible space, visualised conceptually as a quadrilateral, can be systematically spanned to generate a diverse set of solutions.}
  \label{Quadrilateral}
\end{figure}

\begin{figure*}
    \centering
  \includegraphics[width=0.77\textwidth]{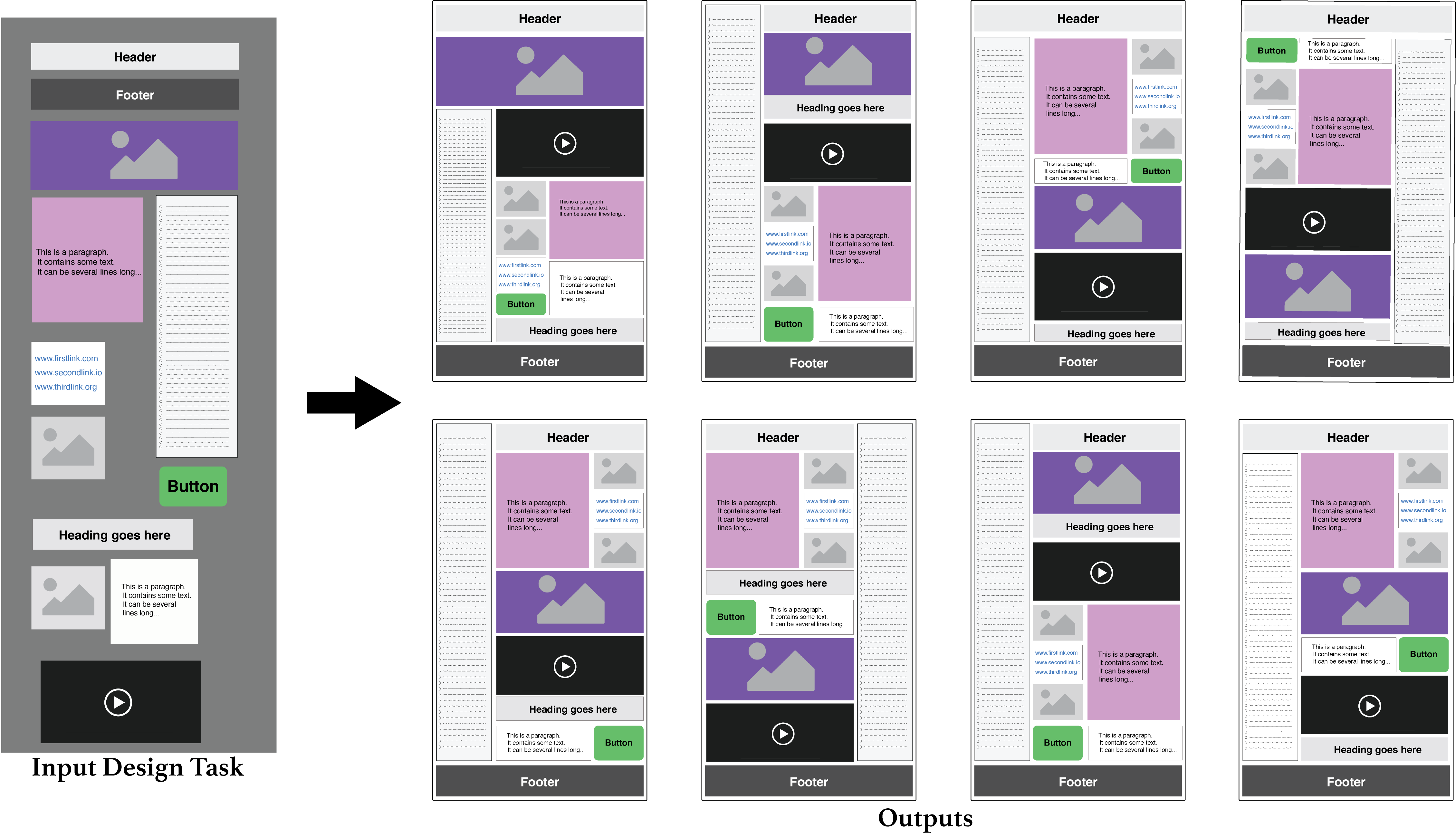}
  \caption{A gallery of diverse resultant grid layouts for a 12-element webpage design, generated by our MILP approach.}
  \label{fig:gallery}
\end{figure*}

We build upon concepts from literature \cite{doi:10.1080/0740817X.2015.1019161, 10.1007/978-3-540-72792-7_22} for systematically spanning the design space. Specifically in our case, the summation of non-negative values of \ElementBeforeElement{}{} and of \ElementAboveElement{}{} constitutes a rigorous two-dimensional \emph{distance metric}. We use this metric to efficiently span the design space by controlling  the amount of dissimilarity required between different solutions. 
Our procedure to obtain a wide variety of diverse solutions is as follows:
First, we solve the optimisation problem while optimising for the criterion of best grid layout.
We obtain the minimal number of grid-lines and designate this as $\mathbb{O}$.
Then we add a constraint that the permissible number of grid lines may not exceed $\mathbb{O}+1$ and we optimise for a rectangular outline.
In case we do not get a rectangular outline, we loosen the constraint of permissible number of grid lines to $\mathbb{O}+2$ and so on.
After getting a rectangular outline, we count the number of cases for adherence to external rectangular hull and set that as a new constraint.
These two constraints in conjunction assure a well-formatted layout with clean grid-lines and rectangular outline.
Next, we focus on generating multiple different solutions.
We solve the optimisation problem again, this time to maximise the value of $\sum\Gamma$.
We designate the resulting objective value as $\Gamma_{max}$.
Then, we minimise $\sum\Gamma$ to find $\Gamma_{min}$.
Similarly, we compute $\Pi_{max}$ and $\Pi_{min}$.
After computing these four boundary values, we consider them as the four cardinal vertices of a conceptual polygon drawn in the 2-D space induced by the interval  $[\Gamma_{min},\Gamma_{max}]$ and $[\Pi_{min},\Pi_{max}]$.
This quadrilateral-shaped feasible space\footnote{We do not argue that the bounds of this representative feasible space are necessarily linear.
The convex quadrilateral with straight edges is only a logical representation and not an argument of linearity.} is illustrated in Figure \ref{Quadrilateral}. 
We seek solutions well spaced across this quadrilateral by enforcing constraints on permissible values of $\sum\Gamma$, $\sum\Pi$.
Successive points in this interval yield solutions that are sufficiently distinct and yet satisfy the alignment requirements discussed earlier.
\autoref{fig:gallery} illustrates diverse solutions for a sample input design.
Systematic spanning of the design space also enables us to search for local improvements in the neighbourhood of a design.

\begin{algorithm}
\caption{: Procedure to generate grid layouts}
\algsetup{indent=2em}
\begin{algorithmic}[1]
    \REQUIRE Data instance involving $n$ elements
    \STATE $\mathbb{F} := \lbrace \rbrace$
    \STATE $\Gamma_{max}, \Gamma_{min} :=$ Extremal values of \ElementAboveElement{}{} by core MILP
    \STATE $\Pi_{max}, \Pi_{min} :=$ Extremal values of \ElementBeforeElement{}{} by core MILP
        \STATE $\epsilon_{min} :=$ Optima value by minimising $\sum \epsilon$
   \STATE $\mathbb{R}_{min}:=$ Optimal number of cases for overall rectangular outline using core MILP
    \STATE Use $\epsilon_{min},\Gamma_{max},\Gamma_{min},\Pi_{max},\Pi_{min},\mathbb{R}_{min}$ to augment core MILP formulation
\WHILE{($\|\mathbb{F}\| < $ Required number of solutions)}
	\FORALL{
    $\Gamma_{val} \in
    [ \Gamma_{min},\Gamma_{max}
    $}
		\FORALL{$
    \Pi_{val} \in
     [\Pi_{min},\Pi_{max} ]
    $}
    		\STATE Enforce $\Gamma_{val}, \Pi_{val}$
            \STATE $f \longleftarrow\ $ Optimal solution of core MILP with current constraints
            \STATE $\mathbb{F} = (\mathbb{F} \cup \{f\})$
              		\STATE Remove $\Gamma_{val}, \Pi_{val}$ constraints
    	\ENDFOR
    \ENDFOR
    \STATE Loosen alignment constraint ($\epsilon_{min}$) by unit value
\ENDWHILE
\ENSURE Set $\mathbb{F}$ of feasible grid layouts
\end{algorithmic}
\label{FullAlgo}
\end{algorithm}

\subsection{Summary}
The full solution procedure, outlined in Algorithm \ref{FullAlgo},  involves repeated application of the core MILP formulation.
Steps 2--5 execute the MILP with different objectives to compute the logical bounds of all governing parameters.
Step 6 restricts the feasible search-space using these bounds. Steps 7--14 enforce suitable constraints. 
Steps 8--10 enable spanning across the design space as identified by \ElementBeforeElement{}{} and \ElementAboveElement{}{}.
Step 11 computes the optimal grid layout solutions. 
Step 14 gradually loosens the alignment constraint if required to ensure that sufficient solutions are generated.

\subsection{Implementation and computational performance}
The model was coded in Python\texttrademark\ version 3.6.
As our solver, we use Gurobi\texttrademark\  on a commodity computer (8-core 64-bit Intel\texttrademark\ i7\texttrademark\ processor 2.8~GHz with 16~GB RAM). 
The optimiser generates well-aligned and properly structured layouts within a very short time.
For example, 5 well-structured layouts are produced within 2 seconds for a 5-element webpage matching the template shown in \autoref{fig:similargrids}. 
For a 12-element webpage design task, illustrated in \autoref{fig:gallery}, the first 5 solutions are generated within 5 seconds;
10--20 further solutions are generated within 30 seconds. 
High performance is attributed to the use of continuous variables in the MILP model. 
Problem sizes are independent of the canvas size; they are $O(n^2)$ polynomial functions of the number of elements. 
The model lends itself to techniques that further improve performance without compromising optimality.

\section{Study 1: Perceived Quality of Grid Layouts }

Using our MILP approach, we can generate grid layouts with controllable levels of grid quality.
To validate our approach, we assessed the quality of resulting layouts through a ratings-based study.
Here, we compared the computed optimality (by optimiser) to the perceived quality (by real users).

\subsection{Method}

\textbf{Participants:}
Recent research has concluded that expert and novice participants alike can differentiate good designs from bad ones \cite{Tanner2019}.
Taking this into account, participation was not restricted to designers.
The study was conducted with 13 participants (7 male, 6 female), consisting of students and researchers at a local university.
Participants were between 22 and 39 years old (Mean 27, SD 5.28).
8 participants had some educational or professional background in (UI) design and/or HCI.
Participation was under informed consent, and the study adhered to European privacy laws (GDPR).

\textbf{Procedure:}
Before the study, we generated a set of 24 layouts for a 12-element case (\autoref{fig:gallery}) in three distinct conditions of optimality (8 designs per condition):
(1) \emph{optimal} (Mean optimality = 98.2\%),
(2) \emph{sub-optimal} (Mean optimality = 61.6\%), and
(3) \emph{far-from-optimal} (Mean optimality = 37.5\%).
All layouts were generated solely using our grid layout objectives.
During the study, each sketch was presented on a display (15 inch Retina MacBook) sequentially.
Presentation order was randomised.

\textbf{Task and Measurement:}
Participants were asked to rate a series of user interface sketches, on a scale of 0 (very bad) to 100 (very good), for their perceived quality.
They were asked to think about well-designed web pages as a baseline.
The participants were free to use their own judgement and were not biased by telling what \emph{well-formed} or \emph{quality} means.

\subsection{Results}
The means for optimal, sub-optimal, and far-from-optimal designs were 72.12 (SD 21.53), 42.66 (SD 20.41), and 20.63 (SD 16.96), respectively.
The effect of optimality on perceived quality was tested with repeated-measures ANOVA.
The effect was found to be significant, \emph{F}(2,12)=213.10, \emph{p}<0.0001.
\autoref{tab:results} summarises these results and compares them with the objected values used in our MILP approach.
Post-hoc Tukey test showed that optimal and sub-optimal conditions had significant difference (\emph{p}<0.0001), optimal and far-from-optimal had significant difference (\emph{p}<0.0001), and sub-optimal and far-from-optimal also had significant difference (\emph{p}<0.0001).

These results corroborate the efficacy of the MILP approach to generate grid layouts with high perceived quality.
It might be noted that the quality rating from participants was consistently lower than the computed optimality.
This could be attributed to the sketch-like appearance of displayed layouts, colouring of elements, and personal preferences (as indicated by large standard deviations).
\begin{table}
\centering
\resizebox{\columnwidth}{!}{%
\begin{tabular}{l|cc|ll}
\multirow{2}{*}{\textbf{Category of instances}} & \multicolumn{2}{c|}{\textbf{Optimality (MILP)}} & \multicolumn{2}{l}{\textbf{Quality Rating}}         \\
                                    & Mean                  & SD                   & \multicolumn{1}{c}{Mean} & \multicolumn{1}{c}{SD} \\ \hline
\textbf{Optimal}                    & 98.2                  & 3.31                   & 72.12                    & 21.53                    \\
\textbf{Sub-Optimal}           & 61.6                  & 13.19                  & 42.66                    & 20.41                    \\
\textbf{Far-from-Optimal}           & 37.5                  & 5.05                   & 20.63                    & 16.96
\end{tabular}%
}
\caption{Mean and standard deviations from MILP and from user ratings. The optimality score had significant impact on user ratings.}
\label{tab:results}
\end{table}

\section{The \toolname\ Design Tool}

\begin{figure}
\centering
\includegraphics[width=0.92\columnwidth]{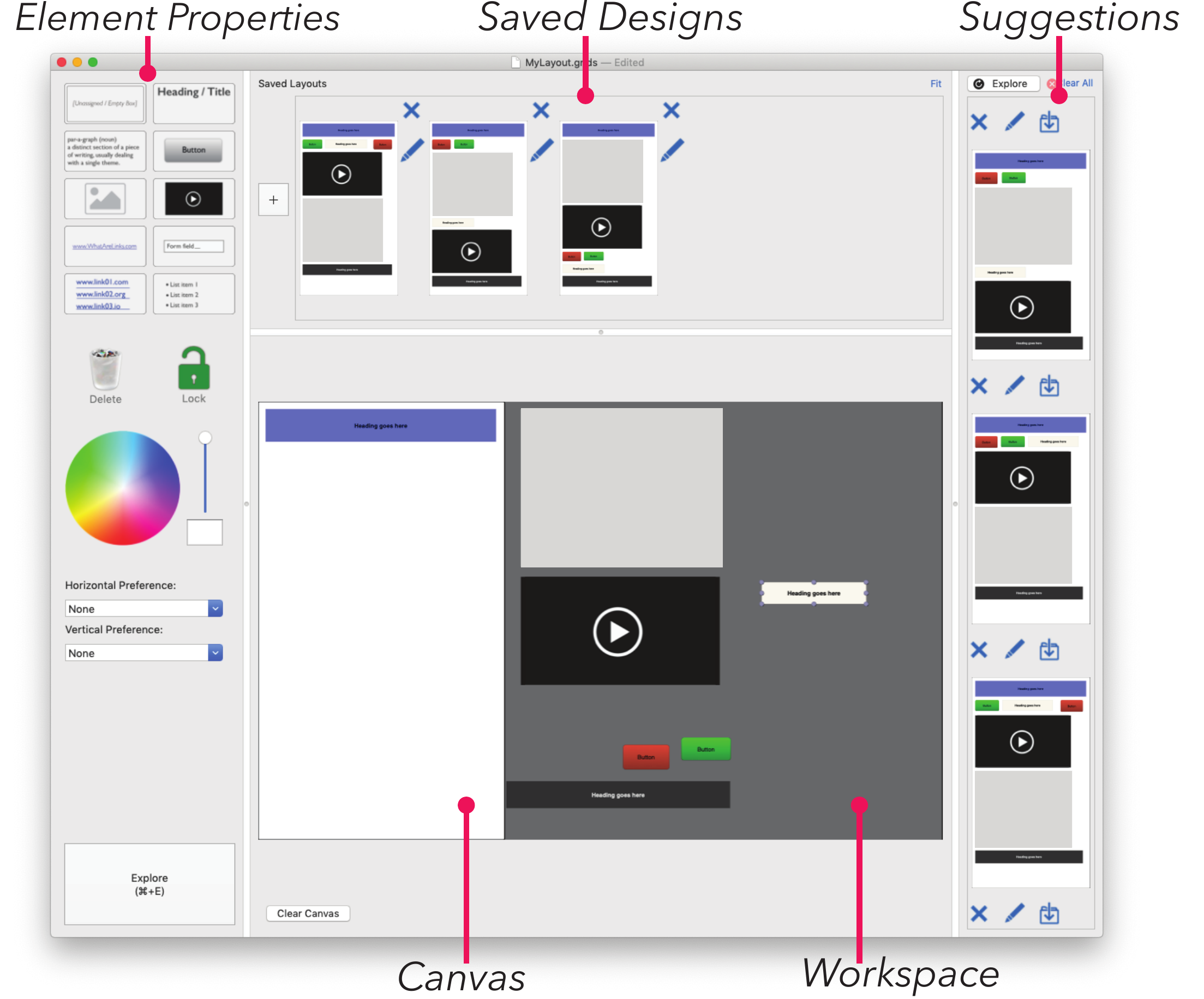}
\caption{The \toolname\ tool presents optimiser-generated layout design suggestions in an interactive example gallery to support wireframing.}
\label{fig:gridstool}
\end{figure}

A key goal in the method has been to support interactive exploration of grid layouts.
To investigate this in practice, we integrated the MILP model and a solver into \toolname,
an interactive tool for wireframing (\autoref{fig:gridstool}). 
As designers sketch layouts, the solver presents them with multiple suggestions in real-time in an example gallery.
This enables designers to quickly detect good solutions and alternatives, and further iterate upon them.
The interface is composed of five regions:
\begin{enumerate}
    \item \textbf{Canvas}: 
    The canvas, sized according to the dimensions of the target UI, is where the designer can concretely place layout elements, defining their size as well as position.
    \item \textbf{Workspace}: 
    Adjacent to the canvas, the workspace is used to specify elements without defining their actual position or size.
    The workspace can also be used as spare working area while moving elements around the canvas.
    \item \textbf{Element Properties Panel}: The left panel is used to specify element properties.
    This includes the element type (heading, paragraph, image, etc.), colour, preferential placement, and whether the element is locked in place.
    The panel also has a button to trigger layout generation and optimisation.
    \item \textbf{Suggestions Panel}: MILP-produced suggestions are displayed in an interactive example gallery \cite{Lee:2010:DIE:1753326.1753667} on the right side.
    As the design task changes, this gallery is refreshed to show updated results.
    Designers can scroll through several suggestions, and edit, save, or delete them.
    \item \textbf{Saved Designs}: At any point, the designer can either save the current canvas, or designs from the suggestions panel.
    These can be viewed, in a gallery, 
    which also serves as a \emph{timeline} showing the evolution of design solutions.
\end{enumerate}

\subsection{Optimiser-Supported Functionalities}
The generated suggestions aid designers during wireframing.
In particular, four key design tasks are supported by GRIDS: 

\textbf{1. Exploring diverse alternatives}:
As the designer specifies layout elements by placing them into the workspace, a diverse set of suggestions is generated and displayed in the \emph{Suggestions} panel.
The designer can select from these, to pick suitable starting points, and can continue iterating over them to achieve a final design.
While similar interactions have been proposed previously (e.g. \cite{todi_sketchplore:_2016}), they could not guarantee coverage of the entire design space.
Typically, designers tend to concentrate their solutions within small parts of the design space \cite{CROSS2004427}.
GRIDS circumvents this issue by facilitating exploration of the entire design space (\autoref{fig:teaser}.1).


\textbf{2. Completing partial solutions}:
When designing a layout, designers often may not have a strong idea on some elements, while some might be pre-determined by the design brief, conventions, or product requirements.
For the optimiser, the main challenge is then to complete the composition of a partial layout.
Completing partial designs has been explored for menu design previously \cite{bailly_menuoptimizer:_2013}, 
where the design of sub-menus was considered.
Our tool supports this for grid layouts. 
Designers can place well-defined elements onto the canvas, while the remaining elements can be specified using the workspace (illustrated in \autoref{fig:teaser}.2).
The optimiser produces results by searching for optimal placements of workspace elements on the partially-filled canvas, and displays a diverse set of complete solutions in the Suggestions panel.
To our knowledge, this has not been shown with previous methods.


\textbf{3. Finding nearby alternatives}:
Previous tools (e.g. \cite{odonovan_designscape:_2015,todi_sketchplore:_2016}) have proposed techniques for finding refinements to a design.
Typically, these focus on providing `fixes' to improve the layout, but not on generating distinct alternatives.
With \toolname, a designer can select any design solution, and request the optimiser for `nearby' alternatives that are close to the input design, but with some variations (illustrated in \autoref{fig:teaser}.3).

\begin{figure}
\centering
\includegraphics[width = 0.99\columnwidth]{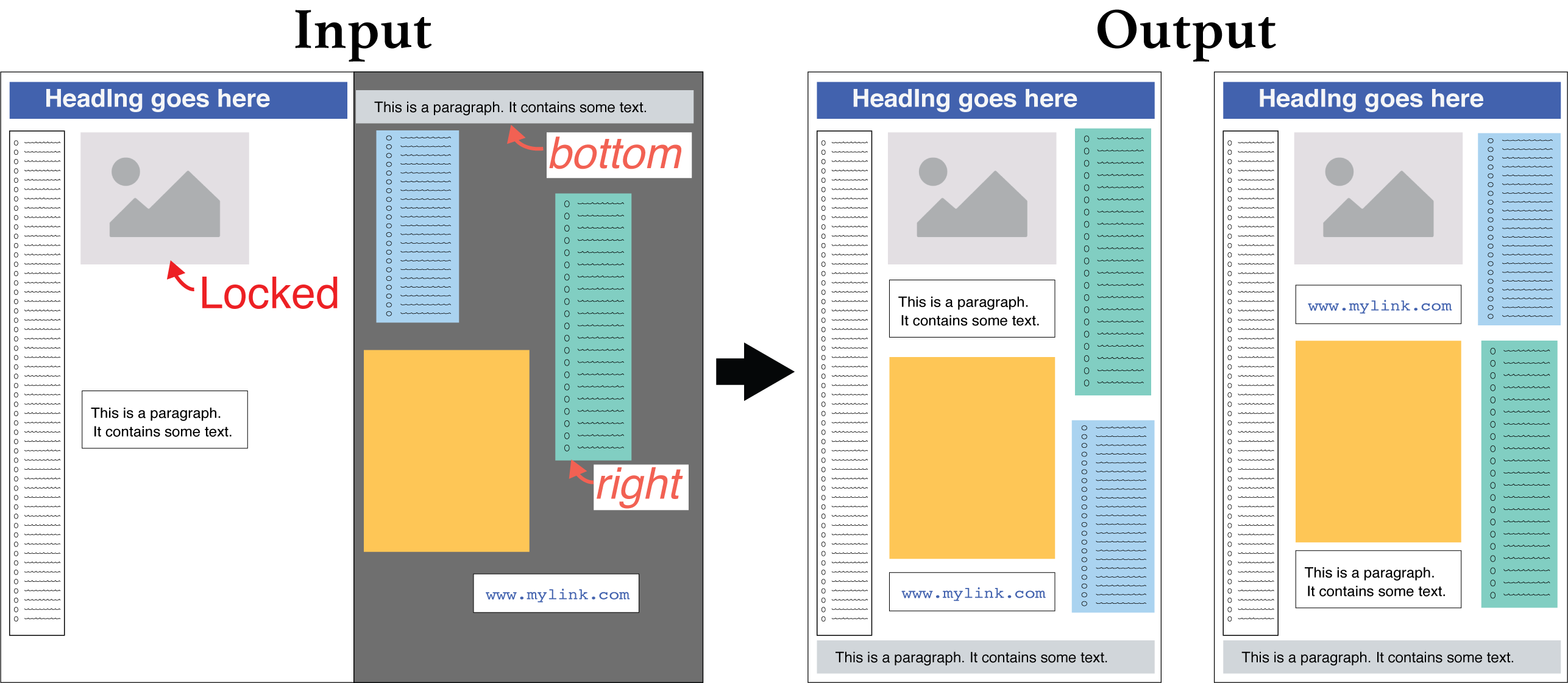}
\caption{Auto-completed solutions within a constrained sub-space can be found by locking elements and specifying placement preferences.}
\label{fig:constrained}
\end{figure}

\textbf{4. Finding solutions in constrained spaces}:
The \emph{Element Properties} panel provides designers with options for constraining the design space (\autoref{fig:constrained}).
Locking has been used previously (e.g. \cite{bailly_menuoptimizer:_2013,todi_sketchplore:_2016}) to restrict some elements from being modified.
By default, canvas elements are flexible, giving the optimiser some freedom in repositioning them to find optimal grid layouts.
Designers can use the \emph{lock} button to fix selected elements on the canvas, and prevent adjustments.
Designers can specify preferential placement of layout elements through drop-down menus.
For instance, they can specify elements as headers or footers by assigning the \textit{vertical preference}, or as left/right sidebars by assigning \textit{horizontal preference}.

\section{Study 2: Evaluation with Designers}
To assess how our approach supports early-stage wireframing, we conducted a study with designers.
Our method follows practices in design and creativity research \cite{dorst2001creativity}.
We aimed for
1) realistic design briefs;
2) a representative sample of professional GUI designers;
3) a mixed-methods approach that gauges both the process and the outcomes of design, including designers' opinions;
and 4) the use of standardised measurements, here for usability and creativity.

\subsection{Method}


\textbf{Participants:} We recruited a total of 16 participants.
Their age ranged from 22 to 34 years (Mean 29.4, SD 3.50).
All had formal design education (Mean 4.53 years, SD 2.32);
15 were professionally employed as designers (Mean 3.97 years, SD 2.58),
while one was still a student.
All had experience with digital design tools (Mean 7.44 years, SD 3.63).
The designers participated under informed consent.
Throughout, the European privacy law (GDPR) was followed.
The participants were compensated with two movie tickets.

\textbf{Design tasks:}
The study consisted of three layout design briefs: (1) a personal blog page with a list of required elements (10 min),
(2) an e-commerce page with freely chosen elements (15 min),
and (3) a free-form task, chosen by the designer, representative of their typical professional design tasks (20 min).
The three tasks ensured variety and sufficient level of complexity to test the system critically.
The third task allowed the designers to freely decide on the type of layout they made,
so all participants could try out the tool in a natural layout design situation they would normally encounter regardless of their background.

\textbf{Setup:} The design tool ran on a MacBook Pro (macOS Mojave 10.14.5).
Participants used the tool with a 24 inch LED display, Apple magic mouse, and Apple wireless keyboard.

\textbf{Procedure:}  First, participants received a short introduction to the tool and a practice task,
where they were asked to complete simple interactions with the tool.
They were allowed to explore the tool until they felt comfortable with it.
The design tasks listed above were then carried out.
For each task, participants were asked to create at least three final designs.
They were free to use the tool features as they wanted.
To be as close to pen-and-paper sketching as possible, the tool, in this study, did not support explicit grouping of elements.

\textbf{Measurements:}
Each of the created designs had metadata on whether it was optimiser-generated, edited by the user, and saved as a final design.
After the tasks, participants filled out the System Usability Scale (SUS) questionnaire (without collaboration) \cite{brooke1996sus},
the Creativity Support Index (CSI) \cite{cherry2014quantifying}.
They also rated usefulness of key features (5-point Likert scale), and answered interview questions related to their experience with the tool and the suggestions, and gave free feedback.

\subsection{Quantitative results}

We analysed the number of designs created by designers, as well as scores from the SUS and CSI questionnaires.

\begin{figure}
    \centering
    \includegraphics[width=0.99\columnwidth]{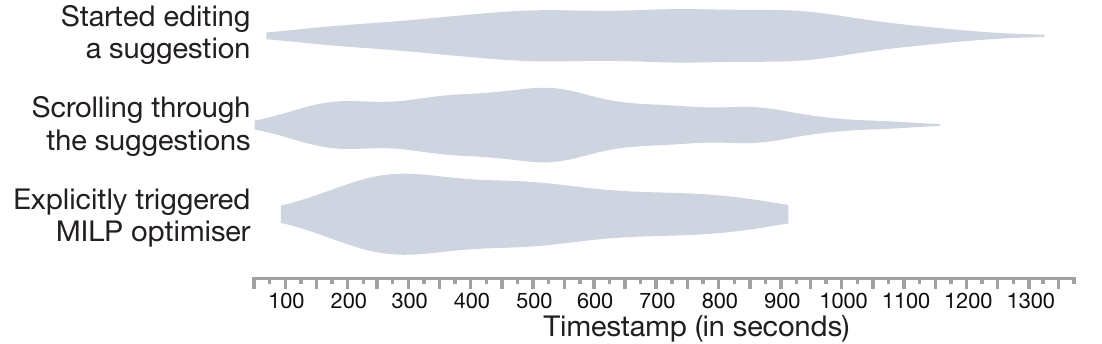}
    \caption{Timeline showing instances of participants triggering the optimiser, scrolling through suggestions, and editing them further.}
    \label{fig:timeline}
\end{figure}

\textbf{Usage of Suggestions:}
When creating design solutions for the three tasks, participants
created a total of 13.75 designs on average (SD = 4.59).
They incorporated suggestions from the optimiser, on average, in 44.17\% (SD = 21.42) of their final designs.
9 out 16 designers used the suggestions in over 50\% of their designs, while only 3 designers sparingly used the suggestions (in <20\% of final designs).
Users triggered the optimiser, and referred to the suggestions at different stages of the design process (early exploration and late refinements), as illustrated in the timeline in \autoref{fig:timeline}.

\textbf{SUS and CSI:}
The mean SUS score was 71.79 (SD 13.95), which according to recommendations can be considered 'good' \cite{bangor2009determining}
The mean CSI score, which focuses on the tool as a whole, was 53.04 (SD 13.58, score range 0-100) and scores for the CSI factors (score range 0-20) were: enjoyment 11.88 (SD 4.29), exploration 12.38 (SD 3.69), expressiveness 9.31 (SD 3.50), immersion 11.5 (SD 3.46), results worth effort 11.88 (SD 3.54).
The mean score is comparable to previous studies using the scale, while exploration was higher than a recent bandit-based system designed explicitly for exploration in moodboard design \cite{koch2019may}.


\textbf{Functionality ratings:} Participants responded being able to \textit{``sketch ideas quickly and freely with the tool''} fairly well (3.81, SD 1.22).
The saved designs timeline was seen as very useful (4.44, SD 0.89),
and the separate canvas and workspace was quite useful (3.75, SD 1.06).
Participants found the exploration feature itself somewhat useful (3.50, SD 1.15), while the usefulness of the suggestions as starting points received mixed feedback 2.8 (SD 1.33).


\subsection{Qualitative results}

Participants found the tool simple and easy to use overall.
8 out of 16 spontaneously commented on ease of use.
Although there were some shortcomings with the tool, because of it being a prototype,
most participants (11 out of 16) commented that liked the idea of interactive layout exploration.
Three participants also pointed out that it is a new idea, \textit{``kind of a new way of thinking about layouts''} (P8).

Participants pointed out that the possible uses for exploration were both in the beginning of the design process to get inspiration, or to fit already existing content on a layout, and later in the process to compare variations of existing layouts.
Exploration was especially seen as useful during early-stage design (14 out of 16), as it could help \textit{``spur creativity and get away from the obvious choices''} (P2).
Four participants commented on fitting required content on a layout e.g. \textit{``sometimes you get requirements of elements you have to fit in the page so the idea of getting different possibilities of where the elements could be is really nice''}.
%
For simple layouts, the exploration works quite well, and participants found the suggestions useful: \textit{``If you have a very simple layout you could take them just as they were''} (P3).
For more complex layouts, the algorithm lacks full support for more complex element hierarchies and groups,
resulting in suggestions being not very relevant to the design at hand.
Participants still said they could get inspiration from those suggestions, but they required more editing:
\textit{``as inspiration to take parts of even though maybe the whole layout isn't useful''} (P3).
It encouraged the designer to think about \textit{``what kind of layout would I design if that element was there''} (P2).



Most participants reported preference for the availability of grid suggestions.
Many (7 out of 16) reported that they usually sketch layout ideas with pen and paper, with most (15 out of 16) also utilising digital tools, mainly Sketch and Adobe applications. Some participants noted that while they could explore their own ideas with these tools, they could not really get new ideas and inspiration in the same way as with this exploration tool.
Exploration is \textit{``a feature that's not available in the industry at this moment''} (P13). Creating rough layouts is \textit{``an important task you have to do all the time in design so ... having such a tool makes sense''} (P3).
Four participants said that they would like to use layout exploration as an integrated part of an existing design tool.
%
Some of the most desired additions to exploration were element grouping, additional exploration dimensions (e.g. colour, accessibility), suggestions based on the layout type, and more settings for elements.
\section{Summary}

We presented a novel mixed integer linear programming (MILP) method for generating grid layouts interactively.
The resulting layouts are well-formed, ensuring proper packing, rectangularity, grouping, and preferential placement. 
We further optimise for grid alignment, a factor correlated with aesthetic perception of layouts \cite{miniukovich2015computation}.
Grid generation at this level of completeness has not been demonstrated with exact methods before.
Study 1 validated the quality of generated layouts.
Owing to the MILP approach, we can \emph{quantify and control} how far grid layouts are from optimum and obtain a well-distributed set across the design space.
Moreover, we have shown this approach to be efficient for interactive use.
In contrast to a recent paper that used black-box optimisation, which throttled at problem sizes of 10 or more elements \cite{todi_sketchplore:_2016},
our model can easily handle such cases.
We demonstrated the feasibility of our approach by implementing it within the \toolname\ tool.
Besides generating controllably diverse sets of alternatives, it enables designers to enforce constraints over the input, and supports features to explore diverse solutions, exploit similar designs, and complete partial layouts.
While our approach could provide immediate benefit to novices by facilitating grid-based layout design, 
we went a step further to evaluate with designers, with professional experience, on the topic.
We received positive feedback from designers about the support for layout exploration,
especially during early stages of design.
We also observed active usage of optimised suggestions in final designs.

\section{Limitations and Future Work}
While our results show that MILP is a promising method to help designers explore grid layouts, 
empirical data pinpoints three opportunities for further improvements that can be built on the approach.
First, Study 2 exposed an issue with element grouping,
which made suggestions less relevant for problems where maintaining original grouping was vital.
While the MILP model supports grouping, 
a designer would need to put effort to express groups via the tool.
Hence, two goals for future work are, 
on the one hand, to develop efficient interaction techniques for specifying grouping, and
on the other,
to extend the MILP model to cover not only contiguity as a visual cue,
but also colour and other commonly-known Gestalt principles \cite{palmer1994rethinking}.

Second, and related, while designers found the approach most useful for early-stage exploration,
it became relatively less useful as the layout became more well-defined.
One goal for future work is to bridge the gap between early-stage grid designs and higher-fidelity prototypes. 
MILP allows flexible and extensible expression of new objectives and constraints, 
however it insists on relaxed or linearised versions of design objectives. 
Some key objectives that need to be identified and formulated for MILP include 
selection and visual search performance, as well as aesthetic objectives, such as colour harmony and clutter.
Finally, the optimisation system presently does not learn.
Our current optimisation model operates on general grid layout principles.
We foresee two possibilities for tuning it using machine learning methods:
(1) If one could detect the designer's ``style'' and whether the designer is exploring or exploiting (e.g., \cite{koch2019may}),
the suggestions could be made even more relevant;
(2) Data-driven approaches could be used to learn layout styles typical to a domain and weigh the parameters of the model to bias it toward that style (e.g, \cite{zheng2019content}).

\section{Open Code}
We support further research efforts by
providing full mathematical formulations in the Supplementary Material, and 
an open code base, with instructions, on our project page: \url{https://userinterfaces.aalto.fi/grids}.

\section{Acknowledgments}
We thank all study participants for their time, and colleagues and reviewers for their helpful comments.
This work has been funded by the European Research Council (ERC) under the European Union's Horizon 2020 research and innovation programme (grant agreement 637991).

\balance{}

\bibliographystyle{SIGCHI-Reference-Format}
\bibliography{references}


\begin{thebibliography}{00}


\ifx \showCODEN    \undefined \def \showCODEN     #1{\unskip}     \fi
\ifx \showDOI      \undefined \def \showDOI       #1{{\tt DOI:}\penalty0{#1}\ }
  \fi
\ifx \showISBNx    \undefined \def \showISBNx     #1{\unskip}     \fi
\ifx \showISBNxiii \undefined \def \showISBNxiii  #1{\unskip}     \fi
\ifx \showISSN     \undefined \def \showISSN      #1{\unskip}     \fi
\ifx \showLCCN     \undefined \def \showLCCN      #1{\unskip}     \fi
\ifx \shownote     \undefined \def \shownote      #1{#1}          \fi
\ifx \showarticletitle \undefined \def \showarticletitle #1{#1}   \fi
\ifx \showURL      \undefined \def \showURL       #1{#1}          \fi

\bibitem{badros_cassowary_2001}
{Greg~J. Badros}, {Alan Borning}, {and} {Peter~J. Stuckey}. 2001.
\newblock \showarticletitle{The {Cassowary} {Linear} {Arithmetic} {Constraint}
  {Solving} {Algorithm}}.
\newblock {\em ACM Trans. Comput.-Hum. Interact.\/} {8}, 4 (Dec. 2001),
  267--306.
\newblock
\showISSN{1073-0516}
\showDOI{%
\url{http://dx.doi.org/10.1145/504704.504705}}


\bibitem{bailly_menuoptimizer:_2013}
{Gilles Bailly}, {Antti Oulasvirta}, {Timo Kötzing}, {and} {Sabrina Hoppe}.
  2013.
\newblock \showarticletitle{{MenuOptimizer}: {Interactive} {Optimization} of
  {Menu} {Systems}}. In {\em Proceedings of the 26th {Annual} {ACM} {Symposium}
  on {User} {Interface} {Software} and {Technology}} {\em ({UIST} '13)}. ACM,
  New York, NY, USA, 331--342.
\newblock
\showISBNx{978-1-4503-2268-3}
\showDOI{%
\url{http://dx.doi.org/10.1145/2501988.2502024}}


\bibitem{balinsky_aesthetically-driven_2009}
{Helen~Y. Balinsky}, {Jonathan~R. Howes}, {and} {Anthony~J. Wiley}. 2009a.
\newblock \showarticletitle{Aesthetically-driven {Layout} {Engine}}. In {\em
  Proceedings of the 9th {ACM} {Symposium} on {Document} {Engineering}} {\em
  ({DocEng} '09)}. ACM, New York, NY, USA, 119--122.
\newblock
\showISBNx{978-1-60558-575-8}
\showDOI{%
\url{http://dx.doi.org/10.1145/1600193.1600219}}


\bibitem{balinsky_aesthetic_2009}
{Helen~Y. Balinsky}, {Anthony~J. Wiley}, {and} {Matthew~C. Roberts}. 2009b.
\newblock \showarticletitle{Aesthetic {Measure} of {Alignment} and
  {Regularity}}. In {\em Proceedings of the 9th {ACM} {Symposium} on {Document}
  {Engineering}} {\em ({DocEng} '09)}. ACM, New York, NY, USA, 56--65.
\newblock
\showISBNx{978-1-60558-575-8}
\showDOI{%
\url{http://dx.doi.org/10.1145/1600193.1600207}}


\bibitem{bangor2009determining}
{Aaron Bangor}, {Philip Kortum}, {and} {James Miller}. 2009.
\newblock \showarticletitle{Determining what individual SUS scores mean: Adding
  an adjective rating scale}.
\newblock {\em Journal of usability studies\/} {4}, 3 (2009), 114--123.
\newblock


\bibitem{bier_snap-dragging_1986}
{Eric~A. Bier} {and} {Maureen~C. Stone}. 1986.
\newblock \showarticletitle{Snap-dragging}. In {\em Proceedings of the 13th
  {Annual} {Conference} on {Computer} {Graphics} and {Interactive}
  {Techniques}} {\em ({SIGGRAPH} '86)}. ACM, New York, NY, USA, 233--240.
\newblock
\showISBNx{978-0-89791-196-2}
\showDOI{%
\url{http://dx.doi.org/10.1145/15922.15912}}


\bibitem{borning_constraints_1997}
{Alan Borning}, {Richard Lin}, {and} {Kim Marriott}. 1997.
\newblock \showarticletitle{Constraints for the {Web}}. In {\em Proceedings of
  the {Fifth} {ACM} {International} {Conference} on {Multimedia}} {\em
  ({MULTIMEDIA} '97)}. ACM, New York, NY, USA, 173--182.
\newblock
\showISBNx{978-0-89791-991-3}
\showDOI{%
\url{http://dx.doi.org/10.1145/266180.266361}}


\bibitem{brooke1996sus}
{John Brooke} {and} {others}. 1996.
\newblock \showarticletitle{SUS-A quick and dirty usability scale}.
\newblock {\em Usability evaluation in industry\/} {189}, 194 (1996), 4--7.
\newblock


\bibitem{chan2017semantically}
{Joel Chan}, {Pao Siangliulue}, {Denisa Qori~McDonald}, {Ruixue Liu}, {Reza
  Moradinezhad}, {Safa Aman}, {Erin~T Solovey}, {Krzysztof~Z Gajos}, {and}
  {Steven~P Dow}. 2017.
\newblock \showarticletitle{Semantically Far Inspirations Considered Harmful?:
  Accounting for Cognitive States in Collaborative Ideation}. In {\em
  Proceedings of the 2017 ACM SIGCHI Conference on Creativity and Cognition}.
  ACM, 93--105.
\newblock


\bibitem{cherry2014quantifying}
{Erin Cherry} {and} {Celine Latulipe}. 2014.
\newblock \showarticletitle{Quantifying the creativity support of digital tools
  through the creativity support index}.
\newblock {\em ACM Transactions on Computer-Human Interaction (TOCHI)\/} {21},
  4 (2014), 21.
\newblock


\bibitem{CHRISTOFIDES199521}
{Nicos Christofides} {and} {Eleni Hadjiconstantinou}. 1995.
\newblock \showarticletitle{An exact algorithm for orthogonal 2-D cutting
  problems using guillotine cuts}.
\newblock {\em European Journal of Operational Research\/} {83}, 1 (1995), 21
  -- 38.
\newblock
\showISSN{0377-2217}
\showDOI{%
\url{http://dx.doi.org/https://doi.org/10.1016/0377-2217(93)E0277-5}}


\bibitem{CROSS2004427}
{Nigel Cross}. 2004.
\newblock \showarticletitle{Expertise in design: an overview}.
\newblock {\em Design Studies\/} {25}, 5 (2004), 427 -- 441.
\newblock
\showISSN{0142-694X}
\showDOI{%
\url{http://dx.doi.org/https://doi.org/10.1016/j.destud.2004.06.002}}
\newblock
\shownote{Expertise in Design.}


\bibitem{10.1007/978-3-540-72792-7_22}
{Emilie Danna}, {Mary Fenelon}, {Zonghao Gu}, {and} {Roland Wunderling}. 2007.
\newblock \showarticletitle{Generating Multiple Solutions for Mixed Integer
  Programming Problems}. In {\em Integer Programming and Combinatorial
  Optimization}, {Matteo Fischetti} {and} {David~P. Williamson} (Eds.).
  Springer Berlin Heidelberg, Berlin, Heidelberg, 280--294.
\newblock


\bibitem{dorst2001creativity}
{Kees Dorst} {and} {Nigel Cross}. 2001.
\newblock \showarticletitle{Creativity in the design process: co-evolution of
  problem--solution}.
\newblock {\em Design studies\/} {22}, 5 (2001), 425--437.
\newblock


\bibitem{drenttel_method_1999}
{William Drenttel} {and} {Jessica Helfand}. 1999.
\newblock Method and system for computer screen layout based on a recombinant
  geometric modular structure.
\newblock   (Aug. 1999).
\newblock
\showURL{%
\url{https://patents.google.com/patent/US7124360B1/en}}


\bibitem{du_plessis_incorporating_2008}
{Mathys~C. du Plessis} {and} {Lynette Barnard}. 2008.
\newblock \showarticletitle{Incorporating {Layout} {Managers} into an
  {Evolutionary} {Programming} {Algorithm} to {Design} {Graphical} {User}
  {Interfaces}}. In {\em Proceedings of the 2008 {Annual} {Research}
  {Conference} of the {South} {African} {Institute} of {Computer} {Scientists}
  and {Information} {Technologists} on {IT} {Research} in {Developing}
  {Countries}: {Riding} the {Wave} of {Technology}} {\em ({SAICSIT} '08)}. ACM,
  New York, NY, USA, 41--47.
\newblock
\showISBNx{978-1-60558-286-3}
\showDOI{%
\url{http://dx.doi.org/10.1145/1456659.1456665}}


\bibitem{freeman-benson_incremental_1990}
{Bjorn~N. Freeman-Benson}, {John Maloney}, {and} {Alan Borning}. 1990.
\newblock \showarticletitle{An {Incremental} {Constraint} {Solver}}.
\newblock {\em Commun. ACM\/} {33}, 1 (Jan. 1990), 54--63.
\newblock
\showISSN{0001-0782}
\showDOI{%
\url{http://dx.doi.org/10.1145/76372.77531}}


\bibitem{frisch_grids_2011}
{Mathias Frisch}, {Sebastian Kleinau}, {Ricardo Langner}, {and} {Raimund
  Dachselt}. 2011.
\newblock \showarticletitle{Grids \& guides: multi-touch layout and alignment
  tools}. In {\em Proceedings of the 2011 annual conference on {Human} factors
  in computing systems - {CHI} '11}. ACM Press, Vancouver, BC, Canada, 1615.
\newblock
\showISBNx{978-1-4503-0228-9}
\showDOI{%
\url{http://dx.doi.org/10.1145/1978942.1979177}}


\bibitem{SUPPLE2}
{Krzysztof Gajos}, {David Christianson}, {Raphael Hoffmann}, {Tal Shaked},
  {Kiera Henning}, {Jing~Jing Long}, {and} {Daniel~S Weld}. 2005.
\newblock \showarticletitle{Fast and robust interface generation for ubiquitous
  applications}. In {\em International Conference on Ubiquitous Computing}.
  Springer, 37--55.
\newblock


\bibitem{gajos_supple:_2004}
{Krzysztof Gajos} {and} {Daniel~S. Weld}. 2004.
\newblock \showarticletitle{{SUPPLE}: {Automatically} {Generating} {User}
  {Interfaces}}. In {\em Proceedings of the 9th {International} {Conference} on
  {Intelligent} {User} {Interfaces}} {\em ({IUI} '04)}. ACM, New York, NY, USA,
  93--100.
\newblock
\showISBNx{978-1-58113-815-3}
\showDOI{%
\url{http://dx.doi.org/10.1145/964442.964461}}


\bibitem{ArnaultSource}
{Krzysztof Gajos} {and} {Daniel~S Weld}. 2005.
\newblock \showarticletitle{Preference elicitation for interface optimization}.
  In {\em Proceedings of the 18th annual ACM symposium on User interface
  software and technology}. ACM, 173--182.
\newblock


\bibitem{gajos2008automatically}
{Krzysztof~Zygmunt Gajos} {and} {Daniel~S Weld}. 2008.
\newblock {\em Automatically generating personalized user interfaces}.
\newblock Citeseer.
\newblock


\bibitem{SUPPLE3}
{Krzysztof~Z Gajos}, {Daniel~S Weld}, {and} {Jacob~O Wobbrock}. 2008.
\newblock \showarticletitle{Decision-Theoretic User Interface Generation.}. In
  {\em AAAI}, Vol.~8. 1532--1536.
\newblock


\bibitem{SUPPLEPLUSPLUS}
{Krzysztof~Z Gajos}, {Jacob~O Wobbrock}, {and} {Daniel~S Weld}. 2007.
\newblock \showarticletitle{Automatically generating user interfaces adapted to
  users' motor and vision capabilities}. In {\em Proceedings of the 20th annual
  ACM symposium on User interface software and technology}. ACM, 231--240.
\newblock


\bibitem{CompareSupple}
{Krzysztof~Z Gajos}, {Jacob~O Wobbrock}, {and} {Daniel~S Weld}. 2008.
\newblock \showarticletitle{Improving the performance of motor-impaired users
  with automatically-generated, ability-based interfaces}. In {\em Proceedings
  of the SIGCHI conference on Human Factors in Computing Systems}. ACM,
  1257--1266.
\newblock


\bibitem{gleicher_briar:_1992}
{Michael Gleicher}. 1992.
\newblock \showarticletitle{Briar: {A} {Constraint}-based {Drawing} {Program}}.
  In {\em Proceedings of the {SIGCHI} {Conference} on {Human} {Factors} in
  {Computing} {Systems}} {\em ({CHI} '92)}. ACM, New York, NY, USA, 661--662.
\newblock
\showISBNx{978-0-89791-513-7}
\showDOI{%
\url{http://dx.doi.org/10.1145/142750.143074}}


\bibitem{HART1995244}
{Stephen~M. Hart} {and} {Liu Yi-Hsin}. 1995.
\newblock \showarticletitle{The application of integer linear programming to
  the implementation of a graphical user interface: a new rectangular packing
  problem}.
\newblock {\em Applied Mathematical Modelling\/} {19}, 4 (1995), 244 -- 254.
\newblock
\showISSN{0307-904X}
\showDOI{%
\url{http://dx.doi.org/https://doi.org/10.1016/0307-904X(94)00033-3}}


\bibitem{hosobe_modular_2001}
{Hiroshi Hosobe}. 2001.
\newblock \showarticletitle{A {Modular} {Geometric} {Constraint} {Solver} for
  {User} {Interface} {Applications}}. In {\em Proceedings of the 14th {Annual}
  {ACM} {Symposium} on {User} {Interface} {Software} and {Technology}} {\em
  ({UIST} '01)}. ACM, New York, NY, USA, 91--100.
\newblock
\showISBNx{978-1-58113-438-4}
\showDOI{%
\url{http://dx.doi.org/10.1145/502348.502362}}


\bibitem{hudson_synergistic_1993}
{Scott~E. Hudson} {and} {Chen-Ning Hsi}. 1993.
\newblock \showarticletitle{A synergistic approach to specifying simple number
  independent layouts by example}. In {\em Proceedings of the {SIGCHI}
  conference on {Human} factors in computing systems - {CHI} '93}. ACM Press,
  Amsterdam, The Netherlands, 285--292.
\newblock
\showISBNx{978-0-89791-575-5}
\showDOI{%
\url{http://dx.doi.org/10.1145/169059.169221}}


\bibitem{hurlburt_grid:_1982}
{Allen Hurlburt}. 1982.
\newblock {\em The {Grid}: {A} {Modular} {System} for the {Design} and
  {Production} of {Newpapers}, {Magazines}, and {Books}\/} (1 edition ed.).
\newblock Wiley, New York; Chichester.
\newblock
\showISBNx{978-0-471-28923-4}


\bibitem{10.1145/882262.882353}
{Charles Jacobs}, {Wilmot Li}, {Evan Schrier}, {David Bargeron}, {and} {David
  Salesin}. 2003.
\newblock \showarticletitle{Adaptive Grid-Based Document Layout}.
\newblock {\em ACM Trans. Graph.\/} {22}, 3 (July 2003), 838–847.
\newblock
\showISSN{0730-0301}
\showDOI{%
\url{http://dx.doi.org/10.1145/882262.882353}}


\bibitem{karrenbauer_improvements_2014}
{Andreas Karrenbauer} {and} {Antti Oulasvirta}. 2014.
\newblock \showarticletitle{Improvements to {Keyboard} {Optimization} with
  {Integer} {Programming}}. In {\em Proceedings of the 27th {Annual} {ACM}
  {Symposium} on {User} {Interface} {Software} and {Technology}} {\em ({UIST}
  '14)}. ACM, New York, NY, USA, 621--626.
\newblock
\showISBNx{978-1-4503-3069-5}
\showDOI{%
\url{http://dx.doi.org/10.1145/2642918.2647382}}


\bibitem{koch2019may}
{Janin Koch}, {Andr{\'e}s Lucero}, {Lena Hegemann}, {and} {Antti Oulasvirta}.
  2019.
\newblock \showarticletitle{May AI?: Design Ideation with Cooperative
  Contextual Bandits}. In {\em Proceedings of the 2019 CHI Conference on Human
  Factors in Computing Systems}. ACM, 633.
\newblock


\bibitem{koffka_principles_2013}
{K. Koffka}. 2013.
\newblock {\em Principles {Of} {Gestalt} {Psychology}}.
\newblock Routledge.
\newblock
\showISBNx{978-1-136-30681-5}
\showDOI{%
\url{http://dx.doi.org/10.4324/9781315009292}}


\bibitem{Lee:2010:DIE:1753326.1753667}
{Brian Lee}, {Savil Srivastava}, {Ranjitha Kumar}, {Ronen Brafman}, {and}
  {Scott~R. Klemmer}. 2010.
\newblock \showarticletitle{Designing with Interactive Example Galleries}. In
  {\em Proceedings of the SIGCHI Conference on Human Factors in Computing
  Systems} {\em (CHI '10)}. ACM, New York, NY, USA, 2257--2266.
\newblock
\showISBNx{978-1-60558-929-9}
\showDOI{%
\url{http://dx.doi.org/10.1145/1753326.1753667}}


\bibitem{DBLP:conf/iclr/LiYHZX19}
{Jianan Li}, {Jimei Yang}, {Aaron Hertzmann}, {Jianming Zhang}, {and} {Tingfa
  Xu}. 2019.
\newblock \showarticletitle{LayoutGAN: Generating Graphic Layouts with
  Wireframe Discriminators}. In {\em 7th International Conference on Learning
  Representations, {ICLR} 2019, New Orleans, LA, USA, May 6-9, 2019}.
\newblock
\showURL{%
\url{https://openreview.net/forum?id=HJxB5sRcFQ}}


\bibitem{LODI2002241}
{Andrea Lodi}, {Silvano Martello}, {and} {Michele Monaci}. 2002.
\newblock \showarticletitle{Two-dimensional packing problems: A survey}.
\newblock {\em European Journal of Operational Research\/} {141}, 2 (2002), 241
  -- 252.
\newblock
\showISSN{0377-2217}
\showDOI{%
\url{http://dx.doi.org/https://doi.org/10.1016/S0377-2217(02)00123-6}}


\bibitem{lok_survey_2001}
{Simon Lok} {and} {Steven Feiner}. 2001.
\newblock {\em A {Survey} of {Automated} {Layout} {Techniques} for
  {Information} {Presentations}}.
\newblock


\bibitem{lok_evaluation_2004}
{Simon Lok}, {Steven Feiner}, {and} {Gary Ngai}. 2004.
\newblock \showarticletitle{Evaluation of {Visual} {Balance} for {Automated}
  {Layout}}. In {\em Proceedings of the 9th {International} {Conference} on
  {Intelligent} {User} {Interfaces}} {\em ({IUI} '04)}. ACM, New York, NY, USA,
  101--108.
\newblock
\showISBNx{978-1-58113-815-3}
\showDOI{%
\url{http://dx.doi.org/10.1145/964442.964462}}


\bibitem{lupton2014thinking}
{Ellen Lupton}. 2014.
\newblock {\em Thinking with type: A critical guide for designers, writers,
  editors, \& students}.
\newblock Chronicle Books.
\newblock


\bibitem{lynch2008web}
{Patrick~J Lynch}. 2008.
\newblock {\em Web style guide}.
\newblock Yale University Press.
\newblock


\bibitem{doi:10.1287/mnsc.44.3.388}
{Silvano Martello} {and} {Daniele Vigo}. 1998.
\newblock \showarticletitle{Exact Solution of the Two-Dimensional Finite Bin
  Packing Problem}.
\newblock {\em Management Science\/} {44}, 3 (1998), 388--399.
\newblock
\showDOI{%
\url{http://dx.doi.org/10.1287/mnsc.44.3.388}}


\bibitem{miniukovich2015computation}
{Aliaksei Miniukovich} {and} {Antonella De~Angeli}. 2015.
\newblock \showarticletitle{Computation of interface aesthetics}. In {\em
  Proceedings of the 33rd Annual ACM Conference on Human Factors in Computing
  Systems}. ACM, 1163--1172.
\newblock


\bibitem{muller-brockmann_grid_1996}
{Josef Müller-Brockmann}. 1996.
\newblock {\em Grid {Systems} in {Graphic} {Design}: {A} {Visual}
  {Communication} {Manual} for {Graphic} {Designers}, {Typographers} and
  {Three} {Dimensional} {Designers}\/} (bilingual edition ed.).
\newblock Niggli, Zürich.
\newblock
\showISBNx{978-3-7212-0145-1}


\bibitem{odonovan_designscape:_2015}
{Peter O'Donovan}, {Aseem Agarwala}, {and} {Aaron Hertzmann}. 2015.
\newblock \showarticletitle{{DesignScape}: {Design} with {Interactive} {Layout}
  {Suggestions}}. In {\em Proceedings of the 33rd {Annual} {ACM} {Conference}
  on {Human} {Factors} in {Computing} {Systems}} {\em ({CHI} '15)}. ACM, New
  York, NY, USA, 1221--1224.
\newblock
\showISBNx{978-1-4503-3145-6}
\showDOI{%
\url{http://dx.doi.org/10.1145/2702123.2702149}}


\bibitem{oney_constraintjs:_2012}
{Stephen Oney}, {Brad Myers}, {and} {Joel Brandt}. 2012.
\newblock \showarticletitle{{ConstraintJS}: {Programming} {Interactive}
  {Behaviors} for the {Web} by {Integrating} {Constraints} and {States}}. In
  {\em Proceedings of the 25th {Annual} {ACM} {Symposium} on {User} {Interface}
  {Software} and {Technology}} {\em ({UIST} '12)}. ACM, New York, NY, USA,
  229--238.
\newblock
\showISBNx{978-1-4503-1580-7}
\showDOI{%
\url{http://dx.doi.org/10.1145/2380116.2380146}}


\bibitem{palmer1994rethinking}
{Stephen Palmer} {and} {Irvin Rock}. 1994.
\newblock \showarticletitle{Rethinking perceptual organization: The role of
  uniform connectedness}.
\newblock {\em Psychonomic bulletin \& review\/} {1}, 1 (1994), 29--55.
\newblock


\bibitem{Adam}
{Seonwook Park}, {Christoph Gebhardt}, {Roman R\"{a}dle}, {Anna~Maria Feit},
  {Hana Vrzakova}, {Niraj~Ramesh Dayama}, {Hui-Shyong Yeo}, {Clemens~N.
  Klokmose}, {Aaron Quigley}, {Antti Oulasvirta}, {and} {Otmar Hilliges}. 2018.
\newblock \showarticletitle{AdaM: Adapting Multi-User Interfaces for
  Collaborative Environments in Real-Time}. In {\em Proceedings of the 2018 CHI
  Conference on Human Factors in Computing Systems} {\em (CHI '18)}. ACM, New
  York, NY, USA, Article 184, 14 pages.
\newblock
\showISBNx{978-1-4503-5620-6}
\showDOI{%
\url{http://dx.doi.org/10.1145/3173574.3173758}}


\bibitem{philippaontology}
{Mothersill Philippa} {and} {Bove~V Michael}. 2018.
\newblock \showarticletitle{An ontology of computational tools for design
  activities}. In {\em Proceedings of DRS}.
\newblock


\bibitem{PISINGER2005154}
{David Pisinger} {and} {Mikkel Sigurd}. 2005.
\newblock \showarticletitle{The two-dimensional bin packing problem with
  variable bin sizes and costs}.
\newblock {\em Discrete Optimization\/} {2}, 2 (2005), 154 -- 167.
\newblock
\showISSN{1572-5286}
\showDOI{%
\url{http://dx.doi.org/https://doi.org/10.1016/j.disopt.2005.01.002}}


\bibitem{Quiroz:2007:HGE:1240866.1241052}
{Juan~C. Quiroz}, {Sergiu~M. Dascalu}, {and} {Sushil~J. Louis}. 2007.
\newblock \showarticletitle{Human Guided Evolution of XUL User Interfaces}. In
  {\em CHI '07 Extended Abstracts on Human Factors in Computing Systems} {\em
  (CHI EA '07)}. ACM, New York, NY, USA, 2621--2626.
\newblock
\showISBNx{978-1-59593-642-4}
\showDOI{%
\url{http://dx.doi.org/10.1145/1240866.1241052}}


\bibitem{sears_aide:_1995}
{Andrew Sears}. 1995.
\newblock \showarticletitle{{AIDE}: {A} {Step} {Toward} {Metric}-based
  {Interface} {Development} {Tools}}. In {\em Proceedings of the 8th {Annual}
  {ACM} {Symposium} on {User} {Interface} and {Software} {Technology}} {\em
  ({UIST} '95)}. ACM, New York, NY, USA, 101--110.
\newblock
\showISBNx{978-0-89791-709-4}
\showDOI{%
\url{http://dx.doi.org/10.1145/215585.215704}}


\bibitem{sutherland_sketchpad:_1963}
{Ivan~E. Sutherland}. 1963.
\newblock \showarticletitle{Sketchpad: {A} {Man}-machine {Graphical}
  {Communication} {System}}. In {\em Proceedings of the {May} 21-23, 1963,
  {Spring} {Joint} {Computer} {Conference}} {\em ({AFIPS} '63 ({Spring}))}.
  ACM, New York, NY, USA, 329--346.
\newblock
\showDOI{%
\url{http://dx.doi.org/10.1145/1461551.1461591}}


\bibitem{Tanner2019}
{Kesler Tanner} {and} {James Landay}. 2019.
\newblock {\em ``I Know It When I See It'': How Experts and Novices Recognize
  Good Design}.
\newblock Springer International Publishing, Cham, 249--266.
\newblock
\showISBNx{978-3-319-97082-0}
\showDOI{%
\url{http://dx.doi.org/10.1007/978-3-319-97082-0_13}}


\bibitem{todi_sketchplore:_2016}
{Kashyap Todi}, {Daryl Weir}, {and} {Antti Oulasvirta}. 2016.
\newblock \showarticletitle{Sketchplore: {Sketch} and {Explore} with a {Layout}
  {Optimiser}}. In {\em Proceedings of the 2016 {ACM} {Conference} on
  {Designing} {Interactive} {Systems}} {\em ({DIS} '16)}. ACM, New York, NY,
  USA, 543--555.
\newblock
\showISBNx{978-1-4503-4031-1}
\showDOI{%
\url{http://dx.doi.org/10.1145/2901790.2901817}}


\bibitem{doi:10.1080/0740817X.2015.1019161}
{Andrew~C. Trapp} {and} {Renata~A. Konrad}. 2015.
\newblock \showarticletitle{Finding diverse optima and near-optima to binary
  integer programs}.
\newblock {\em IIE Transactions\/} {47}, 11 (2015), 1300--1312.
\newblock
\showDOI{%
\url{http://dx.doi.org/10.1080/0740817X.2015.1019161}}


\bibitem{weitzman_automatic_1994}
{L. Weitzman} {and} {Kent Wittenburg}. 1994.
\newblock \showarticletitle{Automatic {Presentation} of {Multimedia}
  {Documents} {Using} {Relational} {Grammars}}. In {\em Proceedings of the
  {Second} {ACM} {International} {Conference} on {Multimedia}} {\em
  ({MULTIMEDIA} '94)}. ACM, New York, NY, USA, 443--451.
\newblock
\showISBNx{978-0-89791-686-8}
\showDOI{%
\url{http://dx.doi.org/10.1145/192593.192718}}


\bibitem{wolsey1998integer}
{Laurence~A Wolsey}. 1998.
\newblock {\em Integer programming}.
\newblock Wiley.
\newblock


\bibitem{xu_global_2014}
{Pengfei Xu}, {Hongbo Fu}, {Takeo Igarashi}, {and} {Chiew-Lan Tai}. 2014.
\newblock \showarticletitle{Global {Beautification} of {Layouts} with
  {Interactive} {Ambiguity} {Resolution}}. In {\em Proceedings of the 27th
  {Annual} {ACM} {Symposium} on {User} {Interface} {Software} and {Technology}}
  {\em ({UIST} '14)}. ACM, New York, NY, USA, 243--252.
\newblock
\showISBNx{978-1-4503-3069-5}
\showDOI{%
\url{http://dx.doi.org/10.1145/2642918.2647398}}


\bibitem{zeidler_constraint_2012}
{Clemens Zeidler}, {Christof Lutteroth}, {and} {Gerald Weber}. 2012a.
\newblock \showarticletitle{Constraint {Solving} for {Beautiful} {User}
  {Interfaces}: {How} {Solving} {Strategies} {Support} {Layout} {Aesthetics}}.
  In {\em Proceedings of the 13th {International} {Conference} of the {NZ}
  {Chapter} of the {ACM}'s {Special} {Interest} {Group} on {Human}-{Computer}
  {Interaction}} {\em ({CHINZ} '12)}. ACM, New York, NY, USA, 72--79.
\newblock
\showISBNx{978-1-4503-1474-9}
\showDOI{%
\url{http://dx.doi.org/10.1145/2379256.2379268}}


\bibitem{zeidler_comparing_2012}
{Clemens Zeidler}, {Johannes Müller}, {Christof Lutteroth}, {and} {Gerald
  Weber}. 2012b.
\newblock \showarticletitle{Comparing the {Usability} of {Grid}-bag and
  {Constraint}-based {Layouts}}. In {\em Proceedings of the 24th {Australian}
  {Computer}-{Human} {Interaction} {Conference}} {\em ({OzCHI} '12)}. ACM, New
  York, NY, USA, 674--682.
\newblock
\showISBNx{978-1-4503-1438-1}
\showDOI{%
\url{http://dx.doi.org/10.1145/2414536.2414638}}


\bibitem{zheng2019content}
{Xinru Zheng}, {Xiaotian Qiao}, {Ying Cao}, {and} {Rynson~WH Lau}. 2019.
\newblock \showarticletitle{Content-aware generative modeling of graphic design
  layouts}.
\newblock {\em ACM Transactions on Graphics (TOG)\/} {38}, 4 (2019), 133.
\newblock


\end{thebibliography}

\end{document}